\documentclass[12pt, draftclsnofoot, onecolumn]{IEEEtran}

\IEEEoverridecommandlockouts

\usepackage{cite}
\usepackage{amsmath,amssymb,amsfonts,amsmath}
\usepackage{algorithmic}
\usepackage{graphicx}
\usepackage{tikz}
\usepackage{tikz-3dplot}
\usepackage{hyperref}
\usetikzlibrary{arrows}
\usetikzlibrary{positioning}   %%<----------------------------
\usepackage{lipsum}
\usetikzlibrary{decorations.markings}
\usetikzlibrary{patterns}
\usetikzlibrary{fit}

\usepackage{textcomp}
\usepackage{xcolor}

\ifCLASSOPTIONcompsoc
    \usepackage[caption=false, font=normalsize, labelfont=sf, textfont=sf]{subfig}
\else
\usepackage[caption=false, font=footnotesize]{subfig}
\fi

\def\BibTeX{{\rm B\kern-.05em{\sc i\kern-.025em b}\kern-.08em
    T\kern-.1667em\lower.7ex\hbox{E}\kern-.125emX}}
\usepackage{mathtools}

\DeclarePairedDelimiter\floor{\lfloor}{\rfloor}

\newcommand{\AxisRotator}[1][rotate=0]{%
    \tikz [x=0.25cm,y=0.60cm,line width=.2ex,-stealth,#1] \draw (0,0) arc (-150:150:1 and 1);%
}
\usepackage{amsthm}
\usepackage{amssymb}
\usepackage[utf8]{inputenc}
\newtheorem{theorem}{Theorem}[]
\newtheorem{lemma}[theorem]{Lemma}

\usepackage{algorithm2e}
\usepackage{amsmath}

\begin{document}
\tikzset{->-/.style={decoration={
			markings,
			mark=at position #1 with {\arrow{>}}},postaction={decorate}}}

\title{A Deep Learning Approach to Location- and Orientation-aided 3D Beam Selection for mmWave Communications

\thanks{This work is supported by the Danish Council for Independent Research, grant no. DFF 8022-00371B.}
}

\author{Sajad~Rezaie,~Elisabeth~de~Carvalho,~and~Carles~Navarro~Manch\'on
\thanks{All the authors are with the Department of Electronic Systems, Aalborg University, Denmark, e-mail: \{sre, 
edc, cnm\}@es.aau.dk}}

% The paper headers
\markboth{}{}

\maketitle

\begin{abstract}
Position-aided beam selection methods have been shown to be an effective approach to achieve high beamforming gain while limiting the overhead and latency of initial access in millimeter wave (mmWave) communications. Most research in the area, however, has focused on vehicular applications, where the orientation of the user terminal (UT) is mostly fixed at each position of the environment. This paper proposes a location- and orientation-based beam selection method to enable context information (CI)-based beam alignment in applications where the UT can take arbitrary orientation at each location. We propose three different network structures, with different amounts of trainable parameters that can be used with different training dataset sizes. A professional 3-dimensional ray tracing tool is used to generate datasets for an IEEE standard indoor scenario. Numerical results show the proposed networks outperform a CI-aided benchmark such as the generalized inverse fingerprinting (GIFP) method as well as hierarchical beam search as a non-CI-based approach. Moreover, compared to the GIFP method, the proposed deep learning-based beam selection shows higher robustness to different line-of-sight blockage probability in the training and test datasets and lower sensitivity to inaccuracies in the position and orientation information.

%Using beam codebooks and deep neural networks, our approach aims to predict a list of the $N$-best beam pairs to communicate with a given UT based only on its position and orientation.
\end{abstract}

%\newpage
\section{Introduction} \label{Intro}
\IEEEPARstart{E}{mergent} services such as virtual and augmented reality or high definition multimedia applications are gaining increasing popularity and will be commonplace in beyond 5G systems. Their large data-rate demands can only be satisfied using large portions of available bandwidth, which millimeter wave (mmWave) bands currently offer \cite{giordani_toward_2020}. Although higher propagation and penetration losses in mmWave bands make mmWave communication a good choice for establishing a network in small cells, those properties, besides the lack of diffraction in mmWave frequencies, make it difficult to establish reliable mmWave links \cite{liu_millimeter-wave_2018}. Multiple-input multiple-output beamforming is an inseparable part of mmWave communication, as it allows for compensating for the higher path loss compared to sub-6 GHz communications. To establish a high-quality directional link, transceivers need to transmit and receive over the direction of the line-of-sight (LOS) path or a powerful non-line-of-sight (NLOS) path in case the LOS is blocked. Thus the selection of appropriate beamforming and combining solutions at mmWave transceivers' antenna arrays is a crucial and challenging task \cite{giordani_initial_2016}. 

Codebook-based beamforming is an attractive solution to reduce the complexity of the beam selection problem as well as the radio-frequency (RF) implementation of beamformers. It considers predefined directional beam configuration sets for precoder and combiner at access point (AP) and user terminal (UT), respectively \cite{xiao_hierarchical_2016}. Analog or hybrid beamforming is usually employed to construct the beams in the codebook. Although the optimal precoder and combiner configurations can be determined by exhaustively searching over all possible combinations of beamformers at the AP and UT, this results in undesirably high overhead and latency. Hierarchical beam search (HBS) is an alternative approach that reduces the search space of the beam alignment procedure by progressively sensing the environment with narrower beam widths based on the counterpart's feedback \cite{xiao_hierarchical_2016, manchon_ping-pong_2017}. However, HBS methods suffer a degradation in accuracy due to low antenna gain and insufficient spatial resolution when using large beam widths, which lead to wrong decisions in the search procedure. %To provide solutions that are optimized for a particular propagation scenario without the need for a highly detailed model of it, machine learning (ML)-based beam alignment methods are proposed. 
To address these downsides, machine learning (ML)-based methods can be used to obtain solutions that are optimized using measurement data obtained at specific deployments without the need for a model of the channel or the propagation scenario.
The authors of \cite{sohrabi_deep_2021-1} propose an ML-based framework that designs adaptive compressed sensing beamformers, which helps estimate the posterior distribution of the angle of arrival (AoA) of the dominant path. Estimates of the AoA and power of the dominant path of users are fed to support vector classifiers and neural networks to perform beam selection in a mmWave communication link in \cite{anton-haro_learning_2019}. However, these methods are designed to perform initial beam alignment in a mmWave environment assuming a single-path channel and are therefore susceptible to performance degradation in multipath conditions. As a new approach to estimates the AoA, a deep-learning based framework is proposed in \cite{polese_deepbeam_2021} that uses the unique patterns in the in-phase-quadrature representation of each beam.

Another avenue to achieve high beamforming accuracy while limiting the overhead and latency of beam alignment is to exploit context information (CI) about the users and/or environment. In CI-based methods, different types of sensors on transceivers or out-band channel measurements are used to acquire CI, at the cost of power and device complexity \cite{kok_using_2017}. As an example of such methods, the inverse fingerprinting (IFP) method in \cite{va_inverse_2018} recommends a candidate beam list based on the location of the user and the history of optimal beam pairs in the training examples. Furthermore, ML-based methods have been widely used to exploit context information and achieve higher accuracy owing to their unparalleled capability in dealing with non-linear problems. Reinforcement learning (RL) and deep neural networks (DNNs) are the two areas of machine learning most frequently employed in mmWave beam alignment \cite{mollel_survey_2021, zhang_reinforcement_2021, xu_3d_2020}. For instance, a RL-based method that uses historical beam management data in a specific environment to reduce the beam training overhead is presented in \cite{shen_design_2021}. However, a large number of interactions with the environment are needed before a RL agent is able to learn the optimal policy \cite{zhu_transfer_2021}. A neural network (NN)-based method is proposed in \cite{alrabeiah_deep_2020} that uses sub-6 GHz channel measurements as input and exploits the spatial correlation between the sub-6 GHz and mmWave links to predict the best beam %index of the codebook 
for data transmission in the mmWave band. For vehicular-to-everything (V2X) application, RADAR signals from joint sensing and communication functionalities at road side units are used for the beamforming of vehicular links in  \cite{liu_radar-assisted_2020}. Also, the authors of \cite{dias_position_2019} propose an ML-based structure that fuses LIDAR data and the position of vehicles to propose a more accurate beam selection method by exploiting knowledge about the obstacles in the environment. As another way of detecting static and mobile obstacles and predicting future link blockage, images from cameras placed in the environment are processed using computer vision and deep learning techniques in \cite{xu_3d_2020, charan_vision-aided_2021, alrabeiah_millimeter_2020}.

Although NN-based beam alignment methods achieve the best performance in beam initialization, beam tracking, and blockage prediction, they require large amounts of training data at each deployment site and for each array configuration \cite{alrabeiah_deep_2020}. As a solution, the transfer learning technique is used in \cite{rezaie_deep_2021} to reuse the knowledge learned by the neural network from a previous environment and array configuration in a new setup. Moreover, one of the key challenges for beam management in beyond 5G systems is device rotation \cite{heng_six_2021}. It becomes more critical as beams in directional codebooks usually have unequal beamwidths and, therefore, the effects of rotation on the received signal strength (RSS) at each beam cannot be fully predicted by the beams' RSS before rotation. In addition, the mentioned machine learning approaches are mainly focused on the V2X application, where the orientation of the UT array can typically be inferred from the vehicle position. The authors of \cite{ali_orientation-assisted_2021} proposed a beam management method using particle filters that exploits the knowledge of changes in the orientation of the user device to track the best beam at the user side. However, this method is limited to beam tracking only at the UT, and it is assumed that the BS has genie-aided knowledge of the best transmit beam. 

\subsection{Contributions}
To address the above-mentioned gaps, we consider the initial beam alignment problem for mmWave communication where the UT can take an arbitrary orientation and position. This study focuses on indoor environments and handset terminals, contrary to many of the works focusing on outdoor V2X communications. We propose a deep learning-based initial beam alignment framework that considers the handset's position and rotation. The proposed solution yields a list of recommended beams that should, later, be sensed by AP and UT. This results in a drastic reduction of the overhead compared to exhaustive search and a significantly improved accuracy compared to HBS. The beam recommendation solution operates using only the position and orientation information of the UT handset as an input. A first version of the method was presented in \cite{rezaie_location-_2020} where an ML-based beam selection method can handle the handset's position and rotation in a 2-dimensional environment. Compared to it, this paper has the following contributions:

\begin{enumerate}
    \item We generated datasets using Altair Feko-Winprop software \cite{noauthor_altair_nodate} as a professional ray-tracing tool. Since UTs can take any arbitrary rotation in a 3-dimensional (3D) environment, it significantly increases the problem's non-linearity. In addition, instead of a uniform linear array (ULA), we use a uniform planar array (UPA) at both sides of the communication links to investigate the effects of rotations in 3D space on the beam alignment of horizontal and vertical beamforming.
    \item We present three different neural networks: single task (ST), multi-task (MT), and extended multi-task (EMT) structures to make accurate beam recommendations with different training dataset sizes. In the MT structure, we solve the problem of the AP and UT beamforming as separate tasks. We extend this idea to also separate vertical and horizontal beamforming in the EMT structure. The MT and EMT structures have significantly fewer trainable parameters than the ST one, resulting in faster performance in the training and running mode at the expense of a slight accuracy loss. 
    \item To enhance the performance of the deep learning-based method, we use the multi-labeling technique to train the NNs. This technique helps the network learn more about the environment and the possible NLOS paths at each location. The effects of the multi-labeling technique on all the three structures DNN-ST, DNN-MT, and DNN-EMT are investigated.
    \item We evaluate the sensitivity of the data-driven-based methods to estimation errors in the location and orientation information by training and evaluating the proposed algorithms with inaccurate CI. In another experiment, we assess the robustness of the proposed deep learning-based methods against mismatch between the channel conditions present in the training and test datasets.  
    \item Our proposed method's performance is compared against several benchmarks. Besides ideal beam alignment, we generalize the inverse fingerprinting method in \cite{va_inverse_2018} to deal also with arbitrary orientation of user handset. In order to compare with a non-CI based method, we extend the deactivation (DEACT) method in \cite{xiao_hierarchical_2016} to perform 3D HBS using UPAs. The proposed deep learning-based beam selection methods significantly outperform the generalized inverse fingerprinting (GIFP)  %as a look-up table type of algorithms 
    and DEACT methods %as a non-context-aware method 
    in terms of accuracy and latency of initial access beam alignment.
\end{enumerate}

\subsection{Organization and Notations}
First, the system model, channel model, and codebook definition are described in Section~\ref{Sec:SysModel}. The general description of the beam alignment problem and two possible approaches for dealing with the problem are given in Section~\ref{Sec:ProbFormulation}. In Section~\ref{Sec:DataDriven}, we present the three proposed structures for the deep neural network. The simulation setup  and numerical results are presented in Section~\ref{Sec:SimResults}. Conclusions are drawn in Section~\ref{Sec:Conc}.

In this paper, $\mathbb{R}$ and $\mathbb{C}$ respectively denote the fields of real and complex numbers.  $\mathcal{A}$ denotes a finite set, with $|\mathcal{A}|$ being its cardinality. Also, $a$, $\boldsymbol{a}$, and $\boldsymbol{A}$ denote a scalar, a vector, and a matrix, respectively, with $a_i$ being the $i$th entry of the column vector $\boldsymbol{a}$, and $A_{i,j}$ is the entry in the $i$th row and $j$th column of the matrix $\boldsymbol{A}$. Transposition and complex transposition of vectors and matrices are respectively represented as $(\cdot)^T$ and $(\cdot)^H$, and  $\otimes$ is the Kronecker product. In addition, $\underset{i, j}{\mathrm{arg\hspace{2pt}max} \hspace{0.5pt} A_{i, j}}$ is the index of the maximum entry of matrix $\boldsymbol{A}$ as a tuple, and $\underset{i, j}{\mathrm{arg\hspace{2pt}sort} \hspace{0.5pt} A_{i, j}}$ denote the list of indices (as tuples) of entries of matrix $\boldsymbol{A}$ sorted in descending order.

\section{System Model}\label{Sec:SysModel}
We consider a downlink communication system consisting of a fixed AP and a mobile UT operating in the mmWave frequency band. Both AP and UT are equipped with UPAs and are placed in a 3D propagation scenario. The received signal at the UT may be written as
\begin{equation}\label{Y_Rec_Sig}
\boldsymbol{y}= \sqrt{P_\mathrm{AP}} \boldsymbol{v}^H \boldsymbol{H} \boldsymbol{u} s + \boldsymbol{v}^H \boldsymbol{n}
\end{equation}
where $\boldsymbol{u}$ and $\boldsymbol{v}$ denote the precoder and combiner at AP and UT, respectively. Also, $P_\mathrm{AP}$, $ s \in \mathbb{C}$, and $\boldsymbol{n} \in \mathbb{C}^{N_\mathrm{UT}}$ , respectively, are the transmission power, the transmitted symbol with unit power, and a complex Gaussian noise vector with zero mean and variance $\sigma_n^2$.

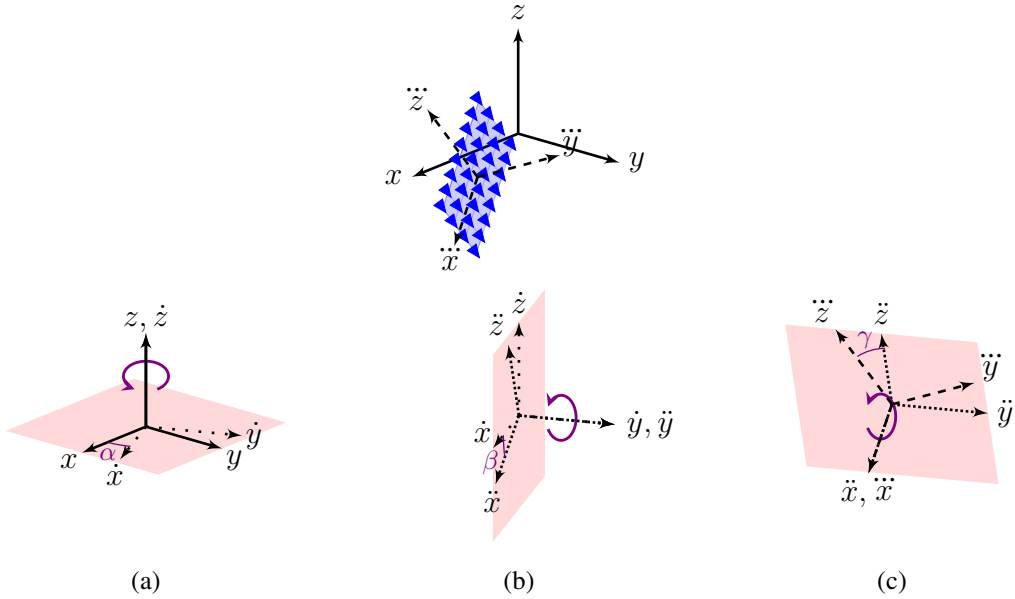
\begin{figure}[t]
\centering
\tdplotsetmaincoords{70}{130}
\begin{tikzpicture}[tdplot_main_coords, scale=0.88]

\draw[>=latex',  ->, line width  = 1pt] (0,0,0) -- ++ (2.5,0,0) node[left =-1pt]{$x$};  % X axis (AP)
\draw[>=latex',  ->, line width  = 1pt] (0,0,0) -- ++ (0,2,0) node[right=-1pt]{$y$};  % Y axis (AP)
\draw[>=latex',  ->, line width  = 1pt] (0,0,0) -- ++ (0,0,1.7) node[above=-1pt]{$z$};  % Z axis (AP)

%\begin{scope}[rotate around z=20] % 15
%    \begin{scope}[rotate around x=25] % 25
%        \begin{scope}[shift={(1.26,0, -0.15)}]. % 7/6cos(30)+0.5sin(30), -7/6sin(30)+0.5cos(30)
%            \begin{scope}[rotate around y=30] % 10
\begin{scope}[shift={(1.55,0.5, -0.15)}]. % 7/6cos(30)+0.5sin(30), -7/6sin(30)+0.5cos(30)            
    \begin{scope}[rotate around z=25] % 15
        \begin{scope}[rotate around y=25] % 25
            \begin{scope}[rotate around x=25] % 10
                \draw[blue, fill=blue, opacity=0.2] (-7/6,0,-0.5) -- ++(7/3,0,0) -- ++(0,0,1) -- ++(-7/3,0,0) -- cycle;
                \draw[blue, opacity=0.2] (1/3-7/6,0,-0.5) -- ++(0,0,1) -- ++(1/3,0,0) -- ++(0,0,-1) -- ++(1/3,0,0) -- ++(0,0,1) -- ++(1/3,0,0) -- ++(0,0,-1)
                -- ++(1/3,0,0) -- ++(0,0,1) -- ++(1/3,0,0) -- ++(0,0,-1)
                -- ++(1/3,0,0) -- ++(0,0,1) ;
                
                \draw[blue, opacity=0.2] (-7/6,0,1/3-0.5) -- ++(7/3,0,0) -- ++(0,0,1/3) -- ++(-7/3,0,0) -- ++(0,0,1/3) -- ++(7/3,0,0) ;
                
                \foreach \i in {0,...,7} {
                    \foreach \j in {0,...,3} {
                        % \filldraw[blue] (\i/3,0,\j/3) circle (1.5pt);
                        \draw[blue, >=triangle 45,  ->, line width  = .1pt] (\i/3-7/6,0,\j/3-0.5) -- (\i/3-7/6,0,\j/3-0.6);
                } 
                }
                
                \draw[>=latex', dashed, ->, line width  = 1pt] (0,0,0) -- ++ (1.003*1.5,0,0) node[below left=-3pt and -7.5pt]{$\dddot{x}$};  % rotated axis x
                \draw[>=latex', dashed, ->, line width  = 1pt] (0,0,0) -- ++ (0,1.003*1.5,0) node[above right=-5pt and -5pt]{$\dddot{y}$};  % rotated axis y
                \draw[>=latex', dashed, ->, line width  = 1pt] (0,0,0) -- ++ (0,0,1.003*1.5) node[above left=-5pt and -5pt]{$\dddot{z}$};  % rotated axis z
            \end{scope}
        \end{scope}
    \end{scope}
\end{scope}
\end{tikzpicture}
%\caption{}
%\label{SystemModel}
%\end{figure}

%%%%%%%%%%%%%%%%%%%
%%%%%%%%%%%%%%%%%%%
%%%%%%%%%%%%%%%%%%%

%\begin{figure}[t]
\centering
\tdplotsetmaincoords{70}{130}
\begin{tikzpicture}[tdplot_main_coords, scale=0.88]

\coordinate (x) at (1.003*1.5,0,0);
\coordinate (y) at (0,1.003*1.5,0);
\coordinate (z) at (0,0,1.003*1.5);

\coordinate (x_d) at (1.003*1.36,1.003*0.63,0);
\coordinate (y_d) at (1.003*-0.63,1.003*1.36,0);
\coordinate (z_d) at (0,0,1.003*1.5);

\coordinate (x_dd) at (1.003*1.23, 1.003*0.57, 1.003*-0.63);
\coordinate (y_dd) at (1.003*-0.63,1.003*1.36,0);
\coordinate (z_dd) at (1.003*0.57, 1.003*0.27, 1.003*1.36);

\coordinate (x_ddd) at (1.003*1.23, 1.003*0.57, 1.003*-0.63);
\coordinate (y_ddd) at (1.003*-0.33, 1.003*1.35, 1.003*0.57);  %(-0.57, 1.12, 0.82);
\coordinate (z_ddd) at (1.003*0.60, 1.003*-0.61, 1.003*1.23);  %(0.63, -0.57, 1.23);

\coordinate (sh1) at (-4,4,0);
\coordinate (sh2) at (-8,8,0);

%%%%%%%%1%%%%%%%%%%
\fill[red!50,opacity=0.3] ($-1*(x) - (y)$) -- ($(x) - (y)$) -- ($(x) + (y)$) -- ($-1*(x) + (y)$) -- ($-1*(x) - (y)$);
\draw[>=latex',  ->, line width  = 1pt] (0,0,0) -- ++ (z) node[below= 5pt]{\AxisRotator[x=0.2cm,y=0.3cm, rotate=90, violet]};  % AxisRotator
\draw[>=latex',  ->, line width  = 1pt] (0,0,0) -- ++ (z) node[above= -4pt]{$z, \dot{z}$};  % Z axis
\draw[>=latex',  ->, line width  = 1pt] (0,0,0) -- ++ (y) node[below right=-4pt and -3pt]{$y$};  % Y axis
\draw[>=latex',  ->, line width  = 1pt] (0,0,0) -- ++ (x) node[below left = -4pt and -3pt]{$x$};  % X axis

\draw[>=latex', loosely dotted, ->, line width  = 1pt] (0,0,0) -- (x_d) node[below left = -3pt and -6pt]{$\dot{x}$};  % X axis
\draw[>=latex', loosely dotted, ->, line width  = 1pt] (0,0,0) -- ++ (y_d) node[right = -3pt]{$\dot{y}$};  % Y axis
\tdplotdrawarc[violet, >=latex', -, line width=0.5pt]{(0,0,0)}{0.9}{0}% tdplot_rotated_coords
{25}{anchor=south, xshift=-4pt, yshift=-10pt}{\textcolor{violet}{\small $\alpha$}};

\node at (0,0,-2.5) {\small (a)};

%%%%%%%%2%%%%%%%%%%

\fill[red!50,opacity=0.3] ($-1*(x_d) - (z_d) + (sh1)$) -- ($(x_d) - (z_d) + (sh1)$) -- ($(x_d) + (z_d)+ (sh1)$) -- ($-1*(x_d) + (z_d)+ (sh1)$) --cycle;
\draw[>=latex', loosely dotted, ->, line width  = 1pt] (sh1) -- ++ (x_d) node[above left=-2pt and -4pt]{$\dot{x}$};  % X axis
\draw[>=latex', loosely dotted, ->, line width  = 1pt] (sh1) -- ++ (y_d) node[right = 0pt]{$\dot{y}, \ddot{y}$};  % Y axis
\draw[>=latex', loosely dotted, ->, line width  = 1pt] (sh1) -- ++ (y_d) node[above right=-12pt and -30pt]{\AxisRotator[x=0.2cm,y=0.3cm, rotate=0, solid, violet]};  % AxisRotator
\draw[>=latex', loosely dotted, ->, line width  = 1pt] (sh1) -- ++ (z_d)node[above=-1pt]{$\dot{z}$};  % Z axis
\draw[>=latex', densely dotted, ->, line width  = 1pt] (sh1) -- ++ (y_dd) node[right=-1pt]{};  % Y axis
\draw[>=latex', densely dotted, ->, line width  = 1pt] (sh1) -- ++ (x_dd) node[below right = -2pt and -9pt]{$\ddot{x}$};  % X axis
\draw[>=latex', densely dotted, ->, line width  = 1pt] (sh1) -- ++ (z_dd) node[above left=-1pt and -3pt]{$\ddot{z}$};  % Z axis
\tdplotgetpolarcoords{1.36}{0.63}{0}
\tdplotsetthetaplanecoords{\tdplotresphi}
\tdplotdrawarc[violet, >=latex', -, line width=0.5pt, tdplot_rotated_coords]{(sh1)}{0.9}{90}% 
{115}{xshift=-5pt, yshift=-6pt}{\textcolor{violet}{\small $\beta$}};

\node at (-4,4,-2.68) {\small (b)};
%%%%%%%%3%%%%%%%%%%

\fill[red!50,opacity=0.3] ($-1*(y_dd) - (z_dd) + (sh2)$) -- ($(y_dd) - (z_dd) + (sh2)$) -- ($(y_dd) + (z_dd)+ (sh2)$) -- ($-1*(y_dd) + (z_dd)+ (sh2)$) --cycle;
\draw[>=latex', densely dotted, ->, line width  = 1pt] (sh2) -- ++ (x_dd) node[below=-2pt]{$\ddot{x}, \dddot{x}$};  % X axis
\draw[>=latex', densely dotted, ->, line width  = 1pt] (sh2) -- ++ (y_dd) node[right=-1pt]{$\ddot{y}$};  % Y axis
\draw[>=latex', densely dotted, ->, line width  = 1pt] (sh2) -- ++ (z_dd) node[above=-1pt]{$\ddot{z}$};  % Z axis
\draw[>=latex', loosely dotted, ->, line width  = 1pt] (sh2) -- ++ (x_dd) node[above right=7pt and -5pt]{\AxisRotator[x=0.2cm,y=0.3cm, rotate=0, solid, violet]}; 

\draw[>=latex', dashed, ->, line width  = 1pt] (sh2) -- ++ (x_ddd) node[above=2pt]{};  % X axis
\draw[>=latex', dashed, ->, line width  = 1pt] (sh2) -- ++ (y_ddd) node[above right=-4pt and -3pt]{$\dddot{y}$};  % Y axis
\draw[>=latex', dashed, ->, line width  = 1pt] (sh2) -- ++ (z_ddd) node[above left=-2pt and -4pt]{$\dddot{z}$};  % Z axis

\tdplotgetpolarcoords{-0.63}{1.36}{0}
\tdplotsetthetaplanecoords{\tdplotresphi}
\tdplotdrawarc[violet, >=latex', -, line width=0.5pt, tdplot_rotated_coords]{(sh2)}{0.9}{-7}%
{-37}{xshift=-2pt, yshift=4pt}{\textcolor{violet}{\small $\gamma$}}:

\node at (-8,8,-2.864) {\small (c)};
\end{tikzpicture}
\caption{By rotating around the $x$, $y$, and $z$ axes, a transition from the global coordinate system ($x \hspace{1pt} y \hspace{1pt} z$) to any arbitrary local coordinate system ($\dddot{x} \hspace{1pt} \dddot{y} \hspace{1pt} \dddot{z}$) is possible. The red rectangles show the rotation planes at three rotation steps. The coordinate systems ($\dot{x} \hspace{1pt} \dot{y} \hspace{1pt} \dot{z}$) and ($\ddot{x} \hspace{1pt} \ddot{y} \hspace{1pt} \ddot{z}$) show the rotated versions of ($x \hspace{1pt} y \hspace{1pt} z$) after the first and second rotation around $z$ and $\dot{y}$ axes, respectively.}
\label{fig:SystemModel}
\end{figure}

We assume that AP and UT UPAs are made respectively of $\{N_\mathrm{AP_H}, N_\mathrm{AP_V}\}$ and $\{N_\mathrm{UT_H}, N_\mathrm{UT_V}\}$ elements in horizontal and vertical dimensions, with the elements spaced half a wavelength apart in both dimensions. The total number of antenna elements at AP and UT is $N_\mathrm{AP} = N_\mathrm{AP_H} N_\mathrm{AP_V}$ and $N_\mathrm{UT} = N_\mathrm{UT_H} N_\mathrm{UT_V}$, respectively. We consider a global coordinate system (GCS) for the environment, and the positions and orientations of AP and UT are defined in the GCS. In addition, both AP and UT have their own local coordinate systems (LCS), shown in Fig. ~\ref{fig:SystemModel}. The LCS is selected such that the UPA is oriented parallel to the $xz$ plane. We assume the AP is placed at fixed position $\boldsymbol{p}_\mathrm{AP} = (x_\mathrm{AP}, y_\mathrm{AP}, z_\mathrm{AP}) \in \mathbb{R}^3$ with fixed angles of antenna arrays $\boldsymbol{\psi}_\mathrm{AP} = (\alpha_\mathrm{AP}, \beta_\mathrm{AP}, \gamma_\mathrm{AP})$ rotated around $z$, $y$, and $x$ axes, respectively. Also, the UT can be placed randomly in the environment at position $\boldsymbol{p}_\mathrm{UT} = (x_\mathrm{UT}, y_\mathrm{UT}, z_\mathrm{UT}) \in \mathbb{R}^3$ with orientation $\boldsymbol{\psi}_\mathrm{UT} = (\alpha_\mathrm{UT}, \beta_\mathrm{UT}, \gamma_\mathrm{UT})$ where the orientation angles are uniformly random in the ranges $\alpha_\mathrm{UT} \in [-\pi, \pi)$, $\beta_\mathrm{UT} \in [-\pi/4, \pi/4)$, and $\gamma_\mathrm{UT} \in [-\pi/4, \pi/4)$. The relation between the GCS, LCS and the orientation angles is illustrated in Fig.~\ref{fig:SystemModel}.

\subsection{Channel Model}
We use Altair WinProp\textsuperscript{TM} software package \cite{noauthor_altair_nodate} to generate channel responses at each UT point in the environment based on the ray-tracing technique. The outputs of the ray-tracing tool such as the angle of departure (AoD), AoA, path gains, etc. for all paths between the AP and each user point are reported as results of the ray-tracing tool. We collect channel responses of UT points at each scene of the environment. The dataset generation steps and assumptions are explained in detail in Section \ref{DataCollection}.

By applying a narrow band channel model on the contributions of one LOS and $L$ NLOS paths, the channel matrix $\boldsymbol{H} \in \mathbb{C}^{N_\mathrm{UT} \times N_\mathrm{AP}}$ between the AP and UT is modeled as
\begin{equation}\label{Eq:Channel}
    \boldsymbol{H} = \sum_{l=0}^{L} \sqrt{\rho_l} \hspace{2pt} e^{j\vartheta_l} \hspace{2pt} \boldsymbol{a}_\mathrm{UT}(\phi^A_{l}, \theta^A_{l}) \hspace{2pt} \boldsymbol{a}_\mathrm{AP}^H(\phi^D_{l},\theta^D_{l})
\end{equation}
where $\rho_l$ and $\vartheta_l$ are the receive power and the phase of the $l$th path, respectively. Also, $\phi^A_{l}$ and $\theta^A_{l}$ denote the azimuth and elevation AoAs with respect to the UT array plane. Likewise, $\phi^D_{l}$ and $\theta^D_{l}$ are the azimuth and elevation AoDs with reference to the AP array plane.
The antenna array response of the UT and AP, respectively $\boldsymbol{a}_\mathrm{UT} \in \mathbb{C}^{N_\mathrm{UT} \times 1}$ and $\boldsymbol{a}_\mathrm{AP} \in \mathbb{C}^{N_\mathrm{AP} \times 1}$, can be written as
\begin{equation}
     \boldsymbol{a}_\mathrm{UT}(\phi^A_{l}, \theta^A_{l}) = \frac{1}{\sqrt{N_\mathrm{UT}}} \boldsymbol{a}_\mathrm{E}(N_\mathrm{UT_V}, \theta^A_{l}) \otimes \boldsymbol{a}_\mathrm{A}(N_\mathrm{UT_H}, \phi^A_{ l}, \theta^A_{l}),
\end{equation}
\begin{equation}
     \boldsymbol{a}_\mathrm{AP}(\phi^D_{l}, \theta^D_{l}) = \frac{1}{\sqrt{N_\mathrm{AP}}} \boldsymbol{a}_\mathrm{E}(N_\mathrm{AP_V},  \theta^D_{l}) \otimes \boldsymbol{a}_\mathrm{A}(N_\mathrm{AP_H}, \phi^D_{ l}, \theta^D_{l})
\end{equation}
where %$\otimes$ is the Kronecker product, and 
$\boldsymbol{a}_\mathrm{A} \in \mathbb{C}^{N \times 1}$ and $\boldsymbol{a}_\mathrm{E} \in \mathbb{C}^{N \times 1}$ are
\begin{equation}\label{eq:a_A}
     \boldsymbol{a}_\mathrm{A}(N, \phi, \theta) =  [1 , e^{j \pi sin(\theta) \hspace{2pt} cos(\phi) }, \dots 
     , e^{ j \pi (N - 1) \hspace{2pt} sin(\theta) \hspace{2pt} cos(\phi) }]^T,
\end{equation}
\begin{equation}\label{eq:a_E}
     \boldsymbol{a}_\mathrm{E}(N, \theta) =  [1 , e^{j \pi cos(\theta)}, \dots 
     , e^{ j \pi (N - 1) \hspace{2pt} cos(\theta) }]^T.
\end{equation}
 
\subsection{Beam Codebook}
We consider analog phased antenna array at both AP and UT. Analog electrically controlled phase shifters are passive devices that can control the signal phase at each array element. We use discrete Fourier transform (DFT)-based codebook to simplify the beamforming procedure where there is one RF chain at each transceiver side. The analog precoder at the AP and the analog combiner at the UT are defined as
\begin{equation}
%\begin{split}
	\boldsymbol{u}_{p,q} = \boldsymbol{a}_\mathrm{AP}(\phi^D_{p}, \theta^D_{q}), \hspace{0.5cm} 
	p\hspace{-2pt}\in\hspace{-2pt}\{1, \dots, N_\mathrm{AP_H}\},\hspace{0.33cm}q\hspace{-2pt}\in\hspace{-2pt}\{1, \dots, N_\mathrm{AP_V}\},
	%  \arccos{(\frac{2q-1-N_{t_z}}{N_{t_z}})}
\end{equation}
\begin{equation}
	\boldsymbol{v}_{m,n} = \boldsymbol{a}_\mathrm{UT}(\phi^A_{m}, \theta^A_{n}), \hspace{0.5cm} m\hspace{-2pt}\in\hspace{-2pt}\{1, \dots, N_\mathrm{UT_H}\},\hspace{0.33cm}n\hspace{-2pt}\in\hspace{-2pt}\{1, \dots, N_\mathrm{UT_V}\}.
%\end{split}
\end{equation}
where $\phi^D_{p}$, $\theta^D_q$, $\phi^A_{m}$ and $\theta^A_n$ are
\begin{equation}
%\begin{split}
    \phi^D_p = \arccos\frac{2p-1-N_\mathrm{AP_H}}{N_\mathrm{AP_H}}, \hspace{5pt} \theta^D_q = \arccos\frac{2q-1-N_\mathrm{AP_V}}{N_\mathrm{AP_V}},
\end{equation}
\begin{equation}
    \phi^A_m = \arccos\frac{2m-1-N_\mathrm{UT_H}}{N_\mathrm{UT_H}}, \hspace{5pt} \theta^A_n = \arccos\frac{2n-1-N_\mathrm{UT_V}}{N_\mathrm{UT_V}}.
%\end{split}
\end{equation}
By mapping all tuples $(p,q)$ and $(m,n)$  into the sets of integers $\{1, 2, \dots, N_\mathrm{AP}\}$ and $\{1, 2, \dots, N_\mathrm{UT}\}$, the sets
$\mathcal{U} = \{\boldsymbol{u}_1, \dots, \boldsymbol{u}_{N_\mathrm{AP}}\}$ and $\mathcal{V} = \{\boldsymbol{v}_1, \dots, \boldsymbol{v}_{N_\mathrm{UT}}\}$ indicate all available analog precoders and combiners at the AP and the UT, respectively. By employing the precoder $\boldsymbol{u}_i$ and combiner $\boldsymbol{v}_j$ at the transceivers, the $(i, j){\text{-th}}$ entry of the RSS matrix, $\boldsymbol{R} \in \mathbb{R}^{N_\mathrm{AP} \times N_\mathrm{UT}}$, can be set as
\begin{equation}\label{Eq:RSS}
R_{i, j} = \Big \lVert \sqrt{P_\mathrm{AP}} \boldsymbol{v}_j^H \boldsymbol{H} \boldsymbol{u}_i s + \boldsymbol{v}_j^H \boldsymbol{n}  \Big \rVert ^2.
\end{equation}
In the beam alignment phase, $s \in \mathbb{C}$ is a pilot symbol with unit power.

\section{Beam Alignment Procedures}\label{Sec:ProbFormulation}
In this section, we discuss the context-aware beam selection and HBS procedures. Afterward, we highlight the similarities and differences between these approaches in terms of required information, training time, and amount of feedback needed between AP and UT.

\subsection{Context-Aware Beam Selection}
The position and orientation of the UT can be used to predict the LOS direction. Although knowledge of the LOS direction is valuable, it is insufficient due to the frequent blockage occurrence in mmWave communications. Besides, the direction and strength of NLOS paths are environment-dependent. Information about the properties of the propagation environment can be obtained from extensive measurements obtained in the specific environment. In Section~\ref{Sec:DataDriven}, we discuss two data-driven methods that use position and orientation information to recommend a list of beam pair candidates.

Consider $\mathcal{B}$ as a set of all possible combinations of precoders and combiners in the transceivers. In the exhaustive search approach, the environment is sensed with all members of $\mathcal{B}$, but this yields unacceptably high overhead \cite{giordani_initial_2016}. Context information from the environment and the transceivers can boost the sensing phase of beam alignment. In particular, context information such as geometrical properties of environment, position and orientation information of transceivers, etc. can be used to reduce the space of sensed precoders and combiners. Thus, in this approach a candidate list $\mathcal{S}$ is proposed based on the context information, where $ | \mathcal{S} | \ll | \mathcal{B} |$.

\begin{figure}
    \centering
    \centering
    \subfloat[Context-aware beam selection\label{fig:Sensing_CABS}]{%
    \scalebox{0.25}{
    \begin{tikzpicture}
        \draw[line width=5pt] (0, 20) node[above = 5pt ]{\Huge \textbf{AP}} -- (0, -4);
        \draw[line width=5pt] (10, 20) node[above = 5pt ]{\Huge \textbf{UT}} -- (10, -4);
        \draw[->, line width=5pt, color=blue] (10, 19) -- (0, 17.5) node[midway, sloped, above = 4pt]{\Huge Req. + Loc + Ori};
        
        \draw[->, line width=5pt, color=blue] (0, 15) -- (10, 13.5) node[midway, sloped, above = 4pt]{\Huge ACK. + beam list $\mathcal{S}$};
        
        \draw[->, line width=5pt, color=red] (0, 13) -- (10, 11.5) node[midway, sloped, above = 4pt]{\Huge beam pair 1};
        
        \draw[->, line width=5pt, color=red] (0, 11.5) -- (10, 10) node[midway, sloped, above = 4pt]{\Huge beam pair 2};
        
        \draw[->, line width=5pt, color=red] (0, 10) -- (10, 8.5) node[midway, sloped, above = 4pt]{\Huge beam pair 3};
        
        \draw[red,fill=red] (5,7.5) circle (.5ex);
        
        \draw[red,fill=red] (5,5.5) circle (.5ex);
        
        \draw[red,fill=red] (5,3.5) circle (.5ex);

        \draw[->, line width=5pt, color=red] (0, 1) -- (10, -0.5) node[midway, sloped, above = 4pt]{\Huge beam pair $|\mathcal{S}|$};
        
        \draw[->, line width=5pt, color=blue] (10, -1.5) -- (0, -3) node[midway, sloped, above = 4pt]{\Huge feedback};
        
        \draw [decorate,xshift=-1cm,yshift=0pt, line width=3pt, color=red]
        (0,0) -- (0,13.5) node [red,rotate=90, midway,yshift=1.5cm] 
        {\Huge AP Beam Training (mmWave)};
        \draw [decorate,xshift=1cm,yshift=0pt, line width=3pt, color=red]
        (10,-1) -- (10,12.5) node [red,rotate=90, midway,yshift=-1.5cm] 
        {\Huge UT Beam Training (mmWave)};
        
        \draw [decorate,xshift=1cm,yshift=0pt, line width=3pt, color=blue]
        (10,19.5) -- (10,13) node [blue,rotate=90, midway,yshift=-1.5cm, text width=7cm] 
        {\Huge \hspace{0.3cm} Training Request \\ \hspace{1cm}(sub-6 GHz)};
        
        \draw [decorate,xshift=1cm,yshift=0pt, line width=3pt, color=blue]
        (10,-1.3) -- (10,-3);
        %\draw[line width=5pt, color=white] (0, -2) -- (0, -2);
        
        \draw[draw=black, fill=green!50] (-1,17.1) rectangle ++(-6,-1.2) node[pos=.5] {\huge Recommender };
        \draw[violet, ->, line width=3pt] (0, 17.3) node [text width=7.5cm] {\vspace{2.8cm} \Huge Loc + Ori} to[out=120,in=+70] (-4, 17.1);
        \draw[violet, ->, line width=3pt] (-4, 15.9) node [text width=-3.5cm] {\vspace{1.2cm} \Huge $\mathcal{S}$} to[out=-70,in=-120] (0, 15.7);
        
        \draw[->, line width=4pt] (-7, 0) -- (-7, -3) node [rotate=90, midway,yshift=1cm]{\Huge Time};
    \end{tikzpicture}
    }} \hspace{3em}%
    %\caption{Context-aware beam selection}
    %\label{fig:Sensing_CABS}
    \subfloat[Hierarchical beam search\label{fig:Sensing_HBS}]{%
    \scalebox{0.25}{
    \begin{tikzpicture}
        \draw[line width=5pt] (0, 19.5) node[above = 5pt ]{\Huge \textbf{AP}} -- (0, -4);
        \draw[line width=5pt] (10, 19.5) node[above = 5pt ]{\Huge \textbf{UT}} -- (10, -4);
        \draw[->, line width=5pt, color=blue] (10, 19) -- (0, 18) node[midway, sloped, above = 4pt]{\Huge Req. };
        
        \draw[->, line width=5pt, color=blue] (0, 17.5) -- (10, 16) node[midway, sloped, above = 4pt]{\Huge ACK.};
        
        \draw[->, line width=5pt, color=red] (0, 16) -- (10, 14.5) node[midway, sloped, above = 4pt]{\Huge UT beam 1};
        
        \draw[->, line width=5pt, color=red] (0, 14.5) -- (10, 13) node[midway, sloped, above = 4pt]{\Huge UT beam 2};
        
        \draw[red,fill=red] (5,13) circle (.5ex);
        \draw[red,fill=red] (5,12.5) circle (.5ex);
        \draw[red,fill=red] (5,12) circle (.5ex);
        
        \draw[->, line width=5pt, color=red] (0, 11) -- (10, 9.5) node[midway, sloped, above = 4pt]{\Huge UT beam $2 \log_2 N_\mathrm{UT}$};
        
        % \draw[->, line width=5pt, color=blue] (10, 9.5) -- (0, 8) node[midway, sloped, above = 4pt]{\Huge feedback};
        
        \draw[->, line width=5pt, color=red] (0, 9) -- (10, 7.5) node[midway, sloped, above = 4pt]{\Huge AP beam 1};
        \draw[->, line width=5pt, color=red] (0, 7.5) -- (10, 6) node[midway, sloped, above = 4pt]{\Huge AP beam 2};
        \draw[->, line width=5pt, color=blue] (10, 5.5) -- (0, 4) node[midway, sloped, above = 4pt]{\Huge feedback};
        
        \draw[red,fill=red] (5,3.5) circle (.5ex);
        \draw[red,fill=red] (5,3) circle (.5ex);
        \draw[red,fill=red] (5,2.5) circle (.5ex);
        
        \draw[->, line width=5pt, color=red] (0, 1.5) -- (10, 0) node[midway, sloped, above = 4pt]{\Huge AP beam $2 \log_2 N_\mathrm{AP} - 1$};
        \draw[->, line width=5pt, color=red] (0, 0) -- (10, -1.5) node[midway, sloped, above = 4pt]{\Huge AP beam $2 \log_2 N_\mathrm{AP}$};
        \draw[->, line width=5pt, color=blue] (10, -2) -- (0, -3.5) node[midway, sloped, above = 4pt]{\Huge feedback};
        
        \draw [decorate,xshift=1cm,yshift=0pt, line width=3pt, color=red]
        (10,12.5) -- (10,15) node [red,rotate=0, midway,xshift=4cm, text width=7cm] 
        {\Huge UT Beam Training \\
        Stage $1$ (mmWave)};
        
        \draw [decorate,xshift=1cm,yshift=0pt, line width=3pt, color=red]
        (10,9) -- (10,11.5) node [red,rotate=0, midway,xshift=5cm, text width=9cm] 
        {\Huge UT Beam Training \\
        Stage $\log_2 N_\mathrm{UT}$ (mmWave)};

        \draw [decorate,xshift=-1cm,yshift=0pt, line width=3pt, color=red]
        (0,6.5) -- (0,9) node [red,rotate=0, midway,xshift=-3.5cm, text width=7cm] 
        {\Huge AP Beam Training \\
        Stage $1$ (mmWave)};
        
        \draw [decorate,xshift=-1cm,yshift=0pt, line width=3pt, color=red]
        (0,-1) -- (0,1.5) node [red,rotate=0, midway,xshift=-4.5cm, text width=9cm] 
        {\Huge \hspace{1cm} AP Beam Training \\
        Stage $\log_2 N_\mathrm{AP}$ (mmWave)};
        
        \draw [decorate,xshift=1cm,yshift=0pt, line width=3pt, color=blue]
        (10,19.5) -- (10,15.5) node [blue,rotate=0, midway,xshift=2.8cm, text width=7cm] 
        {\Huge \hspace{1cm} Training Request \\ \hspace{2cm}(sub-6 GHz)};
        
        %\draw [decorate,decoration={brace,amplitude=10pt},xshift=1cm,yshift=0pt, line width=3pt, color=blue]
        %(10,8) -- (10,9.7);
        
        \draw [decorate,xshift=1cm,yshift=0pt, line width=3pt, color=blue]
        (10,4) -- (10,5.7);
        
        \draw [decorate,xshift=1cm,yshift=0pt, line width=3pt, color=blue]
        (10,-3.5) -- (10,-1.7);
        
        \draw[line width=5pt, color=white] (0, -2) -- (0, -2);
        
        \draw[->, line width=4pt] (17, 0) -- (17, -3) node [rotate=90, midway,yshift=-1cm]{\Huge Time};
    \end{tikzpicture}
    }}
    %\caption{Hierarchical beam search}
    %\label{fig:Sensing_HBS}
    %\end{subfigure}
    \caption{The mmWave beam training procedure for the context-aware beam selection and hierarchical beam search approaches. In the former, the beam list is proposed based on the UT position and orientation.}
    \label{fig:Sensing}
\end{figure}
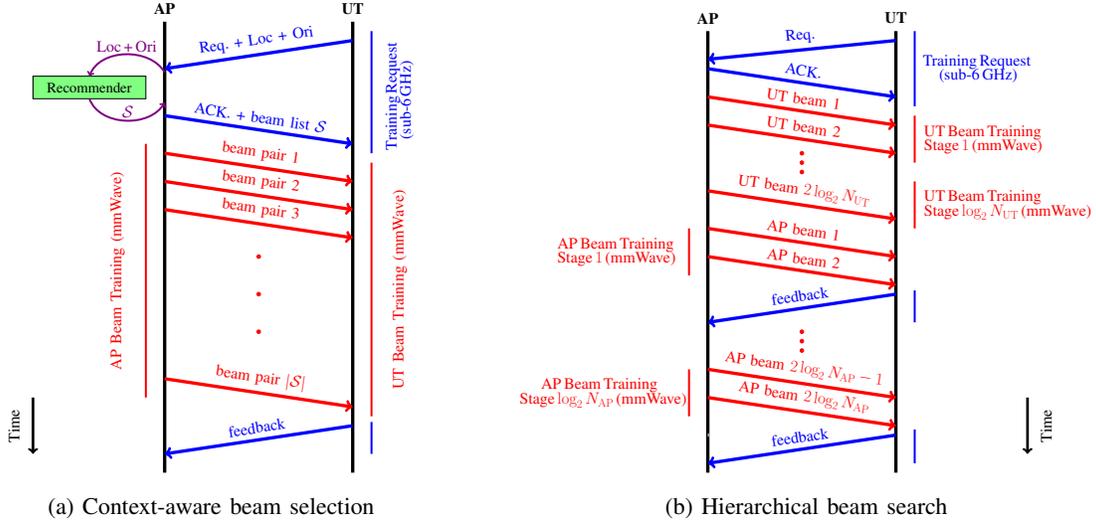

Fig. \ref{fig:Sensing}\subref{fig:Sensing_CABS} shows the beam selection process including the training request and feedback in sub-6 GHz communication links. First, a training request is sent from the UT, including the location and orientation information of the UT array. The AP uses the provided information to recommend a candidate beam list $\mathcal{S}$. Subsequently, the AP shares the beam list $\mathcal{S}$ with the UT. Then, as the AP and UT know their beamformer at each time interval, they start to sense all the beam pairs in $\mathcal{S}$. Then, the UT feeds back the index of the beam pair which resulted the largest RSS. Note that in this work we assume that all control information exchanged over the sub-6 GHz link is conveyed error-free.

\subsection{Hierarchical Beam Search}
In contrast to the context-aware beam selection approach, the hierarchical beam search method does not need extra information about the environment or transceivers. In HBS, the AP and UT sense the environment with wide beams and gradually narrow down the beam-width to find the best beam pairs for transmission \cite{giordani_initial_2016, xiao_hierarchical_2016}. As it is shown in Fig. \ref{fig:Sensing}\subref{fig:Sensing_HBS}, at the first stages AP transmits omnidirectionally, and UT finds the best combiner. Subsequently, AP tries to find its best precoder. At each step of the hierarchical search, it needs feedback from UT to find out the right direction for further search. Thus, $\log_2 N_\mathrm{AP}$ feedbacks are needed to be transmitted in HBS \cite{xiao_hierarchical_2016}. Again, we assume sub-6 GHz feedback to be error-free.

\section{Data-Driven Beam Selection}\label{Sec:DataDriven}
In this section, we discuss the data collection phase as an important step for data-driven based methods and how the labeling can be done for different purposes. Then, we propose the deep-learning based beam selection methods with three novel structures: single-task, multi-task, and extended multi-task. In addition, the motivations for proposing each structure are explained.

\subsection{Dataset structure and construction}\label{DataCollection}
As mentioned in the previous section, the properties of the propagation environment can be extracted from measurements in the specific site. The measurements captured in the particular environment can be collected in a dataset consisting of a set of inputs and outputs.  Here, the position and orientation of measurements are the inputs, and the labels indicating the preferred beams for each input are the outputs.

In the data collection phase, the UT takes an arbitrary location and orientation at each scene of the environment. At each UT position and orientation, the measurements are captured by sensing the environment with all the possible combinations of precoders $\mathcal{U}$ and combiners $\mathcal{V}$ at the transceivers. 
Thus, the entries of $\boldsymbol{R}$ in \eqref{Eq:RSS} are set using the RSS measured using all beam pairs in the codebook. The entries of $\boldsymbol{R}$ are used to set the labels used in the collected datasets. In our work, we consider two types of label encoding options: the one-hot encoding scheme, and the multi-labeling scheme. In the one-hot encoding scheme, the beam pair with the highest RSS is marked in $\boldsymbol{L}$ as the label matrix, where 
\begin{equation}
L^{(1)}_{i, j} = \begin{cases}
		1, & \text{if $(i, j) = \underset{m, n}{\mathrm{arg\hspace{2pt}max}} \hspace{2pt} R_{m, n}$,} \\
		0, & \text{otherwise}.
		\end{cases}
\end{equation}
In this case, in blockage-free situations the beam pair which covers the LOS path is marked\footnote{The presence of measurement noise may impair labeling the LOS direction.}. When blockage occurs, we mark the beam pair covering the strongest NLOS path. 

The one-hot encoding scheme labels only the strongest beam pair for each position and orientation, while all other beam pairs are treated equally. Since the purpose of the recommender in data-driven methods is to propose a list of several candidate beams, it may be useful to, instead, label multiple strong beam pairs as relevant, such that the recommender has information about the M-best beam pairs, rather than just the strongest one. As an example, in a LOS situation such labeling would allow marking both the beam pair capturing the LOS path and a few beam pairs aligned in the direction of strong NLOS paths. This would allow the recommender to, after training, provide a list of alternative candidate beams that can be used in the event that the LOS is blocked. Hence, in the $M$-hot encoding scheme ($M \in \{1, 2, \cdots, N_\mathrm{AP} N_\mathrm{UT}\}$), we mark the $M$ best beam pairs with the $M$ highest RSS in the label matrix $\boldsymbol{L}^{(M)}$ as
\begin{subequations} \label{eq:M}
\begin{align}
& \boldsymbol{r} = \underset{i, j}{\mathrm{arg\hspace{2pt}sort}} \hspace{2pt} (R_{i, j}) ,\\
& \mathcal{T} = \{r_k | k=1, \dots, M\}, \\
& L^{(M)}_{i, j} = 
    \begin{cases}
    1, & \text{if $(i, j) \in  \mathcal{T}$,} \\
    0, & \text{otherwise}.
    \end{cases}
\end{align}
\end{subequations}

Therefore, each sample of the dataset contains the receivers coordinates ($\boldsymbol{p}_\mathrm{UT}$) and orientations ($\boldsymbol{\psi}_\mathrm{UT}$), the RSS matrix ($\boldsymbol{R}$), the type of encoding scheme ($M$), and the label matrix ($\boldsymbol{L}^{(M)}$). We represent the dataset as $\mathbb{D}^{\Xi} = \{(\boldsymbol{x}_m, \boldsymbol{L}^{(M)}_m)\}, m = 1, \cdots, N^{\Xi}$, where $\boldsymbol{x} = [x_\mathrm{UT}, y_\mathrm{UT}, z_\mathrm{UT}, \alpha_\mathrm{UT}, \beta_\mathrm{UT}, \gamma_\mathrm{UT}]$ and $N^{\Xi}$ shows the the dataset size. Indeed, $\boldsymbol{L}^{(M)}$ implicitly holds information about the propagation environment, as it is the post-processed form of $\boldsymbol{R}$. Thus, for given $\boldsymbol{x}$ the dataset contains information about the $M$ strongest beam pairs for transmission, which can be exploited by probabilistic or machine learning based methods \cite{rezaie_location-_2020}.

\subsection{Proposed Deep-Learning Based Methods}
To extract the knowledge about the environment and beam patterns at the transceivers from the training data, probabilistic and machine learning based approaches are the two most popular strategies. In Section \ref{SubSubSec:GIFP}, we discuss a probabilistic beam selection method as a baseline to the proposed methods. Here, we discuss the motivations to use machine learning in beam selection and introduce our machine learning based beam selection methods. 

As static and mobile scatterers in the site may change the propagation properties of the environment by blocking or reflecting paths with high power, beam selection leveraging both location and orientation information becomes a highly nonlinear classification problem. Deep neural networks have shown remarkable achievements in learning complex nonlinear input-output relationships using training data in different applications \cite{goodfellow_deep_2016}. Therefore, we choose deep neural networks as classifiers to predict the best beam pairs for transmission. In the following, we explain the proposed beam selection methods with different network structures and discuss the motivations and assumptions underlying each design.

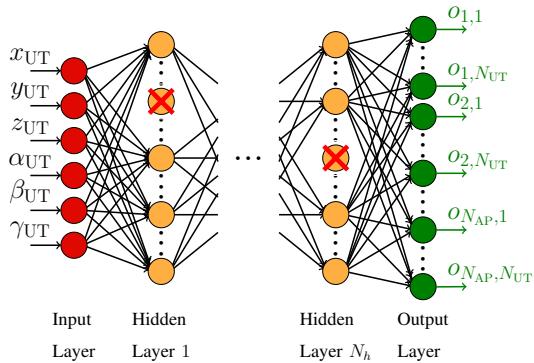
\begin{figure}[t!]
	\centering
	\definecolor{mycolor1}{RGB}{224,9,2} % red
	\definecolor{mycolor2}{RGB}{255,174,66} % orange
	\definecolor{mycolor3}{RGB}{0,120,210} % blue
	\definecolor{mycolor4}{RGB}{0, 128, 0} % Green
	\definecolor{mycolor5}{RGB}{0,191,255} % deep Sky blue
	\definecolor{mycolor6}{RGB}{0,0,139} % dark blue
	\definecolor{mycolor7}{RGB}{255,0,255} % Magenta
	\definecolor{mycolor8}{RGB}{238,130,238} % Violet
	\definecolor{mycolor9}{RGB}{128,0,128} % purple
	\scalebox{.58}{
		\begin{tikzpicture}[cross/.style={path picture={ 
				\draw[red, line width=3pt]
				(path picture bounding box.south east) -- (path picture bounding box.north west) (path picture bounding box.south west) -- (path picture bounding box.north east);
		}}]
		\foreach \i in {-2,...,3} {
			\foreach \h in {-2,...,2} {
				\ifthenelse{\h=1}{}{\draw[black, ->-=0.95, line width=1pt](0.2 , 0.8*\i-0.8*0.5) -- (1.8 ,1.3 * \h);}
			}
		}
		\foreach \i in {-2,...,2} {
			\foreach \h in {-1,...,1} {
				\ifthenelse{\i=1}{}{\draw[black, line width=1pt](2.2 ,1.3 * \i) -- (3.8-0.5 , 1.3 * \h  - 0.5/1.6*1.3*\h +0.5/1.6*1.3*\i);}
			}
		}
		\foreach \i in {-1,...,1} {
			\foreach \h in {-2,...,2} {
				\ifthenelse{\h=0}{}{\draw[black, ->-=0.95, line width=1pt](4.2 +0.5,1.3 * \i + 0.5*1.3*\h/1.6 - 0.5*1.3*\i/1.6) -- (5.8 ,1.3 * \h);}
			}
		}

		\foreach \i in {-2,...,2} {
			\foreach \h in {-2,...,1} {
				\ifthenelse{\i=0}{}{
				\ifthenelse{\h<1}{\draw[black, ->-=0.95, line width=1pt](6.2 ,1.3 * \i) -- (7.8 ,1.3 * \h- 0.35);}{
				\draw[black, ->-=0.95, line width=1pt](6.2 ,1.3 * \i) -- (7.8 ,0.95);}}
			}
		}
		
		\foreach \i in {-2,...,2} {
			\foreach \h in {1,...,2} {
				\ifthenelse{\i=0}{}{\draw[black, ->-=0.95, line width=1pt](6.2 ,1.3 * \i) -- (7.8 ,1.3 * \h+ 0.35);}
			}
		}
		
		\path (3.8,0) -- (4.2,0) node [black, font=\Huge, midway, sloped] {\textbf{$...$}};
		
		\foreach \i in {-2,...,3} {
    		\draw[fill=mycolor1] (0,0.8*\i-0.8*0.5) circle (0.3);  %  Input Layer_i
    	}
		
		\draw[fill=mycolor2] (2,-2.6) circle (0.3);  %  Hidden Layer1
		\path (2,-2.6) -- (2,-1.3) node [black, font=\Huge, midway, sloped] {$...$};
		\draw[fill=mycolor2] (2,-1.3) circle (0.3);  %  Hidden Layer1
		\path (2,-1.3) -- (2,0) node [black, font=\Huge, midway, sloped] {$...$};
		\draw[fill=mycolor2] (2,0) circle (0.3);  %  Hidden Layer1
		\path (2,0) -- (2,1.3) node [black, font=\Huge, midway, sloped] {$...$};
		\draw[fill=mycolor2] (2,1.3) circle (0.3);  %  Hidden Layer1
		\path (2,1.3) -- (2,2.6) node [black, font=\Huge, midway, sloped] {\textbf{$...$}};
		\draw[fill=mycolor2] (2,2.6) circle (0.3);  %  Hidden Layer1
		
		\draw[fill=mycolor2] (6,-2.6) circle (0.3);  %  Hidden Layer5
		\path (6,-2.6) -- (6,-1.3) node [black, font=\Huge, midway, sloped] {$...$};
		\draw[fill=mycolor2] (6,-1.3) circle (0.3);  %  Hidden Layer5
		\path (6,-1.3) -- (6,0) node [black, font=\Huge, midway, sloped] {$...$};
		\draw[fill=mycolor2] (6,0) circle (0.3);  %  Hidden Layer5
		\path (6,0) -- (6,1.3) node [black, font=\Huge, midway, sloped] {$...$};
		\draw[fill=mycolor2] (6,1.3) circle (0.3);  %  Hidden Layer5
		\path (6,1.3) -- (6,2.6) node [black, font=\Huge, midway, sloped] {\textbf{$...$}};
		\draw[fill=mycolor2] (6,2.6) circle (0.3);  %  Hidden Layer5
		
		\draw[fill=mycolor4] (8,-2.95) circle (0.3);  %  Output Layer
		\path (8,-2.95) -- (8,-1.65) node [black, font=\Huge, midway, sloped] {$...$};
		\draw[fill=mycolor4] (8,-1.65) circle (0.3);  %  Output Layer
		\path (8,-1.65) -- (8,-0.35) node [black, font=\Huge, midway, sloped] {$...$};
		\draw[fill=mycolor4] (8,-0.35) circle (0.3);  %  Output Layer
		\path (8,-0.35) -- (8,0.95) node [black, font=\Huge, midway, sloped] {$...$};
		\draw[fill=mycolor4] (8,0.95) circle (0.3);  %  Output Layer
		\draw[fill=mycolor4] (8,1.65) circle (0.3);  %  Output Layer
		\path (8,1.65) -- (8,2.95) node [black, font=\Huge, midway, sloped] {\textbf{$...$}};
		\draw[fill=mycolor4] (8,2.95) circle (0.3);  %  Output Layer
		
		\draw[->, line width=1pt] (-1,-0.8*2-0.8*0.5) node [font=\Large, above] {$\gamma_\mathrm{UT}$} -- (-0.3,-0.8*2-0.8*0.5);
		\draw[->, line width=1pt] (-1,-0.8*1-0.8*0.5) node [font=\Large, above] {$\beta_\mathrm{UT}$} -- (-0.3,-0.8*1-0.8*0.5);
		\draw[->, line width=1pt] (-1,-0.8*0-0.8*0.5) node [font=\Large, above] {$\alpha_\mathrm{UT}$} -- (-0.3,-0.8*0-0.8*0.5);
		\draw[->, line width=1pt] (-1,0.8*1-0.8*0.5) node [font=\Large, above] {$z_\mathrm{UT}$} -- (-0.3,0.8*1-0.8*0.5);
		\draw[->, line width=1pt] (-1,0.8*2-0.8*0.5) node [font=\Large, above] {$y_\mathrm{UT}$} -- (-0.3,0.8*2-0.8*0.5);
		\draw[->, line width=1pt] (-1,0.8*3-0.8*0.5) node [font=\Large, above] {$x_\mathrm{UT}$} -- (-0.3,0.8*3-0.8*0.5);
		
		\node[draw=white,circle,cross,minimum width=0.8 cm] at (2,1.3) {};
		\node[draw=white,circle,cross,minimum width=0.8 cm] at (6,0){};

		\draw[->, line width=1pt, mycolor4] (8.3,-2.955) -- (9,-2.95) node [font=\Large, above right = -2pt and -17pt] {$o_{N_\mathrm{AP}, N_\mathrm{UT}}$};
		\draw[->, line width=1pt, mycolor4] (8.3,-1.65) -- (9,-1.65) node [font=\Large, above right = -2pt and -17pt] {$o_{N_\mathrm{AP}, 1}$};
		\draw[->, line width=1pt, mycolor4] (8.3,-0.35) -- (9,-0.35) node [font=\Large, above right = -2pt and -17pt] {$o_{2, N_\mathrm{UT}}$};
		\draw[->, line width=1pt, mycolor4] (8.3,0.95) -- (9,0.95) node [font=\Large, above right = -2pt and -17pt] {$o_{2, 1}$};
		\draw[->, line width=1pt, mycolor4] (8.3,1.65) -- (9,1.65) node [font=\Large, above right = -2pt and -17pt] {$o_{1, N_\mathrm{UT}}$};
		\draw[->, line width=1pt, mycolor4] (8.3,2.95) -- (9,2.95) node [font=\Large, above right = -2pt and -17pt] {$o_{1, 1}$};

		\node[align=left] at (0,-4.1) {Input \\ Layer };
		\node[align=left] at (2,-4.1) {Hidden \\ Layer $1$};
		\node[align=left] at (6,-4.1) {Hidden \\ Layer $N_{h}$};
		\node[align=left] at (8,-4.1) {Output \\ Layer};
		\end{tikzpicture}}
	\caption{The proposed deep neural network architecture of the beam selection method using position and orientation information. We refer to this network with the term  ``single-task structure'' as $N_\mathrm{AP}*N_\mathrm{UT}$ neurons at its output layer corresponds to all possible beam pairs.}
	\label{DeepLearn-1hot}
\end{figure}
\subsubsection{Single-Task Structure}\label{SubSubSec:ST}
As shown in Fig. \ref{DeepLearn-1hot}, the coordinates and orientations of the UT, $\boldsymbol{x} = [x_\mathrm{UT}, y_\mathrm{UT}, z_\mathrm{UT}, \alpha_\mathrm{UT}, \beta_\mathrm{UT}, \gamma_\mathrm{UT}]$, are fed as inputs to a feed-forward, fully connected, deep neural network. As the AP has fixed position and orientation, there is no need to explicitly feed this information to the NN. There are $N_\mathrm{AP} \times N_\mathrm{UT}$ neurons at the output layer, which correspond to all possible combinations of the $N_\mathrm{AP}$ and $N_\mathrm{UT}$ beams at the AP and UT, respectively. The network has $N_h$ hidden layers with $n_h$ neurons at each hidden layer. Neurons at the $i$th layer of the network, $\boldsymbol{y}^i$, are calculated as a non-linear function of all the neurons at the previous layer, $\boldsymbol{y}^{i-1}$, and the trainable parameters of the $i$th layer:
\begin{equation}
    \boldsymbol{y}^i = \mathit{\sigma} (\hspace{1pt} \boldsymbol{W}^c_i \boldsymbol{y}^{i-1} + \boldsymbol{w}^b_i),  \hspace{0.5cm} 
	i\hspace{-2pt}\in\hspace{-2pt}\{1, \dots, N_{h}+1\}
\end{equation}
where $\sigma$, $\boldsymbol{W}^c_i$ and $\boldsymbol{w}^b_i$ are the activation function, weights and biases of the $i$th layer, respectively. As it is illustrated in Fig. \ref{DeepLearn-1hot}, $\boldsymbol{y}^{0}$ and $\boldsymbol{y}^{N_{h}+1}$ are equal to the input and output vectors, $\boldsymbol{x}$ and $\boldsymbol{o}$, respectively. We define $\boldsymbol{W}_i = \{\boldsymbol{W}^c_i, \boldsymbol{w}^b_i\}$ which includes all the trainable parameters in the $i$th layer. We use $tanh$ function as the non-linear activation function of the hidden layers. By using $softmax$ function as the activation function at the output layer, each network's output represents the predicted probability of each beam pair being the best, meaning providing the largest RSS.
The outputs of the network, $\boldsymbol{o}$, can be expressed as a non-linear function of the input and the weights of different layers:
\begin{equation}
    \boldsymbol{o} = \mathit{f}_{\boldsymbol{W}_o}^{(o)} \Big(\hspace{1pt} \mathit{f}_{\boldsymbol{W}_{N_h}}^{(N_h)}\big(\dots \mathit{f}_{\boldsymbol{W}_1}^{(1)}(\boldsymbol{x} )\big)\Big)
\end{equation}
where $\mathit{f}_{\boldsymbol{W}_o}^{(o)}$ and $\mathit{f}_{\boldsymbol{W}_{i}}^{(i)}, i = 1, \cdots, N_h$ are the non-linear functions of the output layer and the $i$th hidden layer, respectively. With the datasets described in Section \ref{DataCollection}, the network can be trained using a stochastic gradient descent algorithm to learn the most favorable beam pairs for any position in the environment

%The network can learn the good beam pairs for transmission using the single-labeled data, $\boldsymbol{L}^{(1)}$ or multi-labeled data, $\boldsymbol{L}^{(M)}, M \in \{ 2, \cdots, N_\mathrm{AP} N_\mathrm{UT}\}$, which were defined in Section \ref{DataCollection}. 
% With small training datasets, neural networks try to increase the accuracy by memorizing the noise. So, the model fits to the training data perfectly, but the learned model can not be generalized to test samples.
To prevent overfitting, the dropout technique is used as a regularization mechanism at hidden layers \cite{srivastava_dropout_2014}. In this technique, some neurons of the hidden layers are randomly dropped in each training data batch. Dropout helps the network to avoid memorizing training data by effectively using different network structures in the learning process. In Fig. \ref{DeepLearn-1hot}, we show the dropped out neurons in the hidden layers with red cross marks.

Once the model has been trained, it can be used to recommend a set of beam pairs by feeding the UT coordinates and orientation to the network as inputs. A list of recommended beams $\mathcal{S}$ is obtained from the output values $\boldsymbol{o}$, which represent the probability of each beam pair in the codebooks being the best for the input position and orientation. If the sensing phase of the environment is limited by $N_b$ measurements, the $N_b$ beam pairs with highest probabilities are chosen as candidates as follows:
\begin{subequations}
\begin{align}
 & \boldsymbol{g} = \underset{i, j }{\mathrm{arg\hspace{2pt}sort}} \hspace{2pt} (o_{i, j}) ,\\
 & \mathcal{S} = \{g_k | k=1, \dots, N_b\},
\end{align}
\end{subequations}
where $|\mathcal{S}| = N_b$.

\begin{figure}[t!]
	\centering
	\definecolor{mycolor1}{RGB}{224,9,2} % red
	\definecolor{mycolor2}{RGB}{255,174,66} % orange
	\definecolor{mycolor3}{RGB}{0,120,210} % blue
	\definecolor{mycolor4}{RGB}{0, 128, 0} % Green
	\definecolor{mycolor5}{RGB}{0,191,255} % deep Sky blue
	\definecolor{mycolor6}{RGB}{0,0,139} % dark blue
	\definecolor{mycolor7}{RGB}{255,0,255} % Magenta
	\definecolor{mycolor8}{RGB}{238,130,238} % Violet
	\definecolor{mycolor9}{RGB}{128,0,128} % purple
	\scalebox{.58}{
		\begin{tikzpicture}[cross/.style={path picture={ 
				\draw[red, line width=3pt]
				(path picture bounding box.south east) -- (path picture bounding box.north west) (path picture bounding box.south west) -- (path picture bounding box.north east);
		}}]
		\foreach \i in {-2,...,3} {
			\foreach \h in {-2,...,2} {
				\ifthenelse{\h=1}{}{\draw[black, ->-=0.95, line width=1pt](0.2 , 0.8*\i-0.8*0.5) -- (1.8 ,1.3 * \h);}
			}
		}
		
		\foreach \i in {-2,...,2} {
			\foreach \h in {-1,...,1} {
				\ifthenelse{\i=1}{}{\draw[black, line width=1pt](2.2 ,1.3 * \i) -- (3.8-0.5 ,1.3 * \h  - 0.5/1.6*1.3*\h + 0.5/1.6*1.3*\i);}
			}
		}
		\foreach \i in {-1,...,1} {
			\foreach \h in {-2,...,2} {
				\ifthenelse{\h=0}{}{\draw[black, ->-=0.95, line width=1pt](4.2 +0.5,1.3 * \i + 0.5*1.3*\h/1.6 - 0.5*1.3*\i/1.6) -- (5.8 ,1.3 * \h);}
			}
		}
		
		\foreach \i in {-2,...,2} {
			\foreach \h in {0,...,1} {
				\ifthenelse{\i=0}{}{\draw[black, ->-=0.95, line width=1pt](6.2 ,1.3 * \i) -- (7.8 ,1.3 * \h + 0.8);}
			}
		}
		\foreach \i in {-2,...,2} {
			\foreach \h in {0,...,-1} {
				\ifthenelse{\i=0}{}{\draw[black, ->-=0.95, line width=1pt](6.2 ,1.3 * \i) -- (7.8 ,1.3 * \h -0.8);}
			}
		}
		
		\path (3.8,0) -- (4.2,0) node [black, font=\Huge, midway, sloped] {\textbf{$...$}};
		
		\foreach \i in {-2,...,3} {
    		\draw[fill=mycolor1] (0,0.8*\i-0.8*0.5) circle (0.3);  %  Input Layer_i
    	}
		
		\draw[fill=mycolor2] (2,-2.6) circle (0.3);  %  Hidden Layer1
		\path (2,-2.6) -- (2,-1.3) node [black, font=\Huge, midway, sloped] {$...$};
		\draw[fill=mycolor2] (2,-1.3) circle (0.3);  %  Hidden Layer1
		\path (2,-1.3) -- (2,0) node [black, font=\Huge, midway, sloped] {$...$};
		\draw[fill=mycolor2] (2,0) circle (0.3);  %  Hidden Layer1
		\path (2,0) -- (2,1.3) node [black, font=\Huge, midway, sloped] {$...$};
		\draw[fill=mycolor2] (2,1.3) circle (0.3);  %  Hidden Layer1
		\path (2,1.3) -- (2,2.6) node [black, font=\Huge, midway, sloped] {\textbf{$...$}};
		\draw[fill=mycolor2] (2,2.6) circle (0.3);  %  Hidden Layer1
		
		\draw[fill=mycolor2] (6,-2.6) circle (0.3);  %  Hidden Layer5
		\path (6,-2.6) -- (6,-1.3) node [black, font=\Huge, midway, sloped] {$...$};
		\draw[fill=mycolor2] (6,-1.3) circle (0.3);  %  Hidden Layer5
		\path (6,-1.3) -- (6,0) node [black, font=\Huge, midway, sloped] {$...$};
		\draw[fill=mycolor2] (6,0) circle (0.3);  %  Hidden Layer5
		\path (6,0) -- (6,1.3) node [black, font=\Huge, midway, sloped] {$...$};
		\draw[fill=mycolor2] (6,1.3) circle (0.3);  %  Hidden Layer5
		\path (6,1.3) -- (6,2.6) node [black, font=\Huge, midway, sloped] {\textbf{$...$}};
		\draw[fill=mycolor2] (6,2.6) circle (0.3);  %  Hidden Layer5
		
		\draw[fill=mycolor7] (8,-2.1) circle (0.3);  %  Output Layer
		\path (8,-2.1) -- (8,-0.8) node [black, font=\Huge, midway, sloped] {$...$};
		\draw[fill=mycolor7] (8,-0.8) circle (0.3);  %  Output Layer
		\draw[fill=mycolor3] (8,0.8) circle (0.3);  %  Output Layer
		\path (8,0.8) -- (8,2.1) node [black, font=\Huge, midway, sloped] {\textbf{$...$}};
		\draw[fill=mycolor3] (8,2.1) circle (0.3);  %  Output Layer
		
		\draw[->, line width=1pt] (-1,-0.8*2-0.8*0.5) node [font=\Large, above] {$\gamma_\mathrm{UT}$} -- (-0.3,-0.8*2-0.8*0.5);
		\draw[->, line width=1pt] (-1,-0.8*1-0.8*0.5) node [font=\Large, above] {$\beta_\mathrm{UT}$} -- (-0.3,-0.8*1-0.8*0.5);
		\draw[->, line width=1pt] (-1,-0.8*0-0.8*0.5) node [font=\Large, above] {$\alpha_\mathrm{UT}$} -- (-0.3,-0.8*0-0.8*0.5);
		\draw[->, line width=1pt] (-1,0.8*1-0.8*0.5) node [font=\Large, above] {$z_\mathrm{UT}$} -- (-0.3,0.8*1-0.8*0.5);
		\draw[->, line width=1pt] (-1,0.8*2-0.8*0.5) node [font=\Large, above] {$y_\mathrm{UT}$} -- (-0.3,0.8*2-0.8*0.5);
        
		\draw[->, line width=1pt] (-1,0.8*3-0.8*0.5) node [font=\Large, above] {$x_\mathrm{UT}$} -- (-0.3,0.8*3-0.8*0.5);
		\node[draw=white,circle,cross,minimum width=0.8 cm] at (2,1.3) {};
		\node[draw=white,circle,cross,minimum width=0.8 cm] at (6,0){};
		
		\draw[->, line width=1pt, mycolor7] (8.3,-2.1) -- (9.5,-2.1) node [font=\Large, above right = -2pt and -37pt] {$o^r_{N_\mathrm{UT}}$};
		\draw[->, line width=1pt, mycolor7] (8.3,-0.8) -- (9.5,-0.8) node [font=\Large, above right = -2pt and -37pt] {$o^r_{1}$};
		\draw[->, line width=1pt, mycolor3] (8.3,0.8) -- (9.5,0.8) node  [font=\Large, above right=-2pt and -37pt] {$o^t_{N_\mathrm{AP}}$};
		\draw[->, line width=1pt, mycolor3] (8.3,2.1) -- (9.5,2.1) node [font=\Large, above right=-2pt and -37pt] {$o^t_{1}$};
		
		\node[align=left] at (0,-4.1) {Input \\ Layer };
		\node[align=left] at (2,-4.1) {Hidden \\ Layer 1};
		\node[align=left] at (6,-4.1) {Hidden \\ Layer $N_{h}$};
		\node[align=left] at (8,-4.1) {Output \\ Layer};

	    \coordinate (dm1) at (10,-3);
        \coordinate (dm2) at (10,3);
        \node[rectangle, draw, mycolor4, fill=mycolor4!5, very thick, minimum width=1cm] [fit = (dm1) (dm2)] (bx4) {};
		\node[align=center,font=\large,rotate=90] at (bx4.center) {Outer product + Flatten};
		
		\draw[->, line width=1pt, mycolor4] (10.5,-2.95) -- (11.5,-2.95) node [font=\Large, above right = -2pt and -17pt] {$o_{N_\mathrm{AP}, N_\mathrm{UT}}$};
		\draw[->, line width=1pt, mycolor4] (10.5,-1.65) -- (11.5,-1.65) node [font=\Large, above right = -2pt and -17pt] {$o_{N_\mathrm{AP}, 1}$};
		\draw[->, line width=1pt, mycolor4] (10.5,-0.35) -- (11.5,-0.35) node [font=\Large, above right = -2pt and -17pt] {$o_{2, N_\mathrm{UT}}$};
		\draw[->, line width=1pt, mycolor4] (10.5,0.95) -- (11.5,0.95) node [font=\Large, above right = -2pt and -17pt] {$o_{2, 1}$};
		\draw[->, line width=1pt, mycolor4] (10.5,1.65) -- (11.5,1.65) node [font=\Large, above right = -2pt and -17pt] {$o_{1, N_\mathrm{UT}}$};
		\draw[->, line width=1pt, mycolor4] (10.5,2.95) -- (11.5,2.95) node [font=\Large, above right = -2pt and -17pt] {$o_{1, 1}$};
		
		\path (10.8,1.65) -- (10.8,2.95) node [black, font=\Huge, midway, sloped, mycolor4] {\textbf{$...$}};
		\path (10.8,-0.35) -- (10.8,0.95) node [black, font=\Huge, midway, sloped, mycolor4] {\textbf{$...$}};
		\path (10.8,-1.65) -- (10.8,-0.35) node [black, font=\Huge, midway, sloped, mycolor4] {\textbf{$...$}};
		\path (10.8,-2.95) -- (10.8,-1.65) node [black, font=\Huge, midway, sloped, mycolor4] {\textbf{$...$}};
		\end{tikzpicture}}
		
	\caption{The multi-task structure of the proposed deep neural network. There are two separate sets of neurons at the output layer, corresponding to beams at AP and UT. So, there are $N_\mathrm{AP} + N_\mathrm{UT}$ neurons at the out layer, leading to less trainable parameters than the single-task structure.}
	\label{DeepLearn-MultiTask-1}
\end{figure}
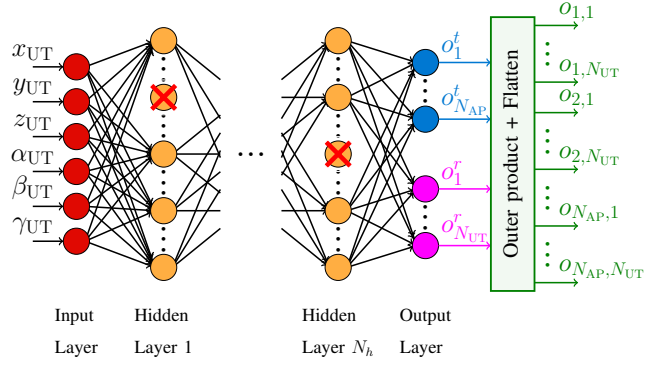
\subsubsection{Multi-Task Structure}\label{SubSubSec:MT}
Although the single-task structure of the neural network can propose good beam candidates for a given UT position and orientation, it needs large training datasets to be trained well due to the large number of trainable weights it contains. By inspecting the number of trainable parameters at each layer, we see that there are $7 n_h$, $(n_h + 1) n_h$, and $(n_h+1) N_\mathrm{AP} N_\mathrm{UT}$ trainable weights, respectively, at the first hidden layer, other hidden layers, and the output layer. For large number of antenna elements, it is likely that $N_\mathrm{AP} N_\mathrm{UT} \gg n_h$. In this case, the output layer contains most of the trainable parameters. By changing the structure of the output layer, we can decrease the number of trainable weights significantly. 

We propose a network structure that attempts to solve the AP and UT beamformer selection as separate problems or tasks.
As shown in Fig. \ref{DeepLearn-MultiTask-1}, we consider two separate sets of neurons at the output layer, with different tasks for each branch: beam selection at the AP and beam selection at the UT. The two sets of neurons, $\boldsymbol{o}^t = [o^t_1, \dots,  o^t_{N_\mathrm{AP}}]^T$ and $\boldsymbol{o}^r = [o^r_1, \dots,  o^r_{N_\mathrm{UT}}]^T$, show the probability of each beam being the best at AP and UT for a given UT position and orientation. As the network aims to propose good beam pairs to be sensed, we synthetically combine the probabilities of beams at the AP and the UT to construct probabilities of all possible beam pairs. Following this simplifying assumption, we use the outer product to calculate the probability of each beam pair being the best, $o_{i, j}$, i.e.
\begin{equation} \label{eq:outer}
    o_{i,j} = o^t_i \hspace{1pt} o^r_j, \hspace{0.25cm} i \in \{1, \dots, N_\mathrm{AP}\}, \hspace{0.25cm} j \in \{1, \dots, N_\mathrm{UT}\}.
\end{equation}
%where $o^t_i$ and $o^r_j$ denote the probability of $i$th and $j{\text{-th}}$ beam being the best at AP and UT, respectively.
As the outer product of two vectors $\boldsymbol{o}^t$ and $\boldsymbol{o}^r$ is a matrix by size $N_\mathrm{AP} \times N_\mathrm{UT}$, we flatten the result of the outer product to obtain a vector with $N_\mathrm{AP} N_\mathrm{UT}$ entries, $\boldsymbol{o}$ \footnote{The operation ``outer product + flatten'' is equivalent to the Kronecker product, i.e. $\boldsymbol{o} = \boldsymbol{o}^t \otimes \boldsymbol{o}^r$ .}. As it is clear in \eqref{eq:outer}, there is no trainable parameter in the construction of probabilities of beam pairs. As the output layer has $N_\mathrm{AP}+N_\mathrm{UT}$ neurons, the number of trainable weights at the output layer is $(n_h+1) (N_\mathrm{AP}+N_\mathrm{UT})$. In practical scenarios where the training dataset is limited, the multi-task structured network can be trained better than the single-task structured one, as the former has far fewer trainable parameters.

Although combining the output neurons of different tasks is not common in multi-task neural network structures in other applications, it is necessary to combine them here as AP and UT do beamforming simultaneously. In other words, the network should recommend a list of beam pairs for sensing as a candidate list. The following lemma shows the relation between the loss function of the proposed multi-task structure calculated before and after the outer product.

\begin{lemma} % \textbf{Lemma 1.}
In case the categorical cross entropy is used as the loss function, $\mathcal{L}$, of the multi-task structured network, the loss function of the form $\mathcal{L}(\boldsymbol{o}, \boldsymbol{L}^{(M)})$ is equivalent to $\mathcal{L}(\boldsymbol{o}^t, \boldsymbol{l}^t)+\mathcal{L}(\boldsymbol{o}^r, \boldsymbol{l}^r)$, where $l^t_i = \sum_{j=1}^{N_\mathrm{UT}} L^{(M)}_{i,j}$ and $l^r_j = \sum_{i=1}^{N_\mathrm{AP}} L^{(M)}_{i,j}$.

% Proof.
\begin{proof}
\begin{subequations}
\begin{align}
    \mathcal{L}(\boldsymbol{o}, \boldsymbol{L}^{(M)}) &= - \sum_{i=1}^{N_\mathrm{AP}} \sum_{j=1}^{N_\mathrm{UT}} L^{(M)}_{i,j} \log(o_{i,j}) \\
    %&= - \sum_{i=1}^{N_\mathrm{AP}} \sum_{j=1}^{N_\mathrm{UT}} L^{(M)}_{i,j} \log(o^t_i o^r_j) \\
    &= - \sum_{i=1}^{N_\mathrm{AP}} \sum_{j=1}^{N_\mathrm{UT}} L^{(M)}_{i,j} \Big( \log(o^t_i) +  \log(o^r_j) \Big)\\
    %&= - \sum_{i=1}^{N_\mathrm{AP}} \sum_{j=1}^{N_\mathrm{UT}} L^{(M)}_{i,j} \log(o^t_i) - \sum_{j=1}^{N_\mathrm{UT}} \sum_{i=1}^{N_\mathrm{AP}} L^{(M)}_{i,j} \log(o^r_j) \\
    &= - \sum_{i=1}^{N_\mathrm{AP}} l^t_i \log(o^t_i) - \sum_{j=1}^{N_\mathrm{UT}} l^r_j \log(o^r_j) \\
    &= \mathcal{L}(\boldsymbol{o}^t, \boldsymbol{l}^t)+\mathcal{L}(\boldsymbol{o}^r, \boldsymbol{l}^r)
\end{align}
\end{subequations}
\end{proof}

\end{lemma}

The lemma highlights that even though the outer product operation is not typical in neural networks structures, the training of the multi-task structured network is similar to the standard training procedure in multi-task applications. In the single-label case $M = 1$, training the network with label $L_{i,j}=1$ after the outer product is equal to training with labels $l^t_{i}=1$ and $l^r_{j}=1$ at the two output branches before the outer product. In the case $M>1$, however, AP and UT beams that appear in multiple non-zero entries in the beam-pair label matrix $\boldsymbol{L}^{(M)}$ are given larger weight in the equivalent AP and UT beam label vectors $\boldsymbol{l}^t$ and $\boldsymbol{l}^r$.

\subsubsection{Extended Multi-Task Structure}\label{SubSubSec:EMT}
To further decrease the number of trainable weights, we consider the assumption that the horizontal and vertical beamforming are independent at both AP and UT. Based on this simplifying assumption, we propose an extended version of the multi-task structured network, which is shown in Fig. \ref{DeepLearn-MultiTask-2}. There are 4 separate sets of neurons for horizontal and vertical beamforming at AP and UT. Similar to the previous structure, we use the outer product to obtain the probabilities of being the best for all possible beams at transceivers, i.e., 
\begin{equation} \label{eq:outer_2}
    o^t_{i} = o^{t^A}_m \hspace{1pt} o^{t^E}_n, \hspace{0.1cm} i \in \{1, \dots, N_\mathrm{AP}\}, \hspace{0.05cm} m=\floor*{\frac{i}{N_\mathrm{AP_H}}}, \hspace{0.05cm} n= i\bmod N_\mathrm{AP_H},
\end{equation}
\begin{equation}
    o^r_{j} = o^{r^A}_p \hspace{1pt} o^{r^E}_q, \hspace{0.1cm} j \in \{1, \dots, N_\mathrm{UT}\}, \hspace{0.05cm} p=\floor*{\frac{j}{N_\mathrm{UT_H}}}, \hspace{0.05cm} q= j\bmod N_\mathrm{UT_H},
\end{equation}
where $o^{t^A}_m$, $o^{t^E}_n$, $o^{r^A}_p$, and $o^{r^E}_q$ denote the probability of $m$th, $n$th, $p$th, and $q$th beams being the best at horizontal and vertical beamforming at AP and UT, respectively. As an example, neuron $o^{t^A}_m$ denotes the output probability for the AP beams in the $m$th azimuth angle of the codebook, while $o^{t^E}_n$ represents the probability for the AP beams pointing at the $n$th elevation angle. As in the new structure the output layer includes $N_\mathrm{AP_H}+N_\mathrm{AP_V}+N_\mathrm{UT_H}+N_\mathrm{UT_V}$ neurons, the number of trainable parameters in the output layer decreases to $(n_h+1) (N_\mathrm{AP_H}+N_\mathrm{AP_V}+N_\mathrm{UT_H}+N_\mathrm{UT_V})$. Such a reduction is advantageous when there amount of training data is limited.

\begin{figure}[t!]
	\centering
	\definecolor{mycolor1}{RGB}{224,9,2} % red
	\definecolor{mycolor2}{RGB}{255,174,66} % orange
	\definecolor{mycolor3}{RGB}{0,120,210} % blue
	\definecolor{mycolor4}{RGB}{0, 128, 0} % Green
	\definecolor{mycolor5}{RGB}{0,191,255} % deep Sky blue
	\definecolor{mycolor6}{RGB}{0,0,139} % dark blue
	\definecolor{mycolor7}{RGB}{255,0,255} % Magenta
	\definecolor{mycolor8}{RGB}{180,100,180} % Violet
	\definecolor{mycolor9}{RGB}{128,0,128} % purple
	\scalebox{.58}{
		\begin{tikzpicture}[cross/.style={path picture={ 
				\draw[red, line width=3pt]
				(path picture bounding box.south east) -- (path picture bounding box.north west) (path picture bounding box.south west) -- (path picture bounding box.north east);
		}}]
		\foreach \i in {-2,...,3} {
			\foreach \h in {-2,...,2} {
				\ifthenelse{\h=1}{}{\draw[black, ->-=0.95, line width=1pt](0.2 , 0.8*\i-0.8*0.5) -- (1.8 ,1.3 * \h);}
			}
		}
		\foreach \i in {-2,...,3} {
			\foreach \h in {-2,...,2} {
				\ifthenelse{\h=1}{}{\draw[black, ->-=0.95, line width=1pt](0.2 , 0.8*\i-0.8*0.5) -- (1.8 ,1.3 * \h);}
			}
		}
		
		\foreach \i in {-2,...,2} {
			\foreach \h in {-1,...,1} {
				\ifthenelse{\i=1}{}{\draw[black, line width=1pt](2.2 ,1.3 * \i) -- (3.8-0.5 ,1.3 * \h  - 0.5/1.6*1.3*\h + 0.5/1.6*1.3\i);}
			}
		}
		\foreach \i in {-1,...,1} {
			\foreach \h in {-2,...,2} {
				\ifthenelse{\h=0}{}{\draw[black, ->-=0.95, line width=1pt](4.2 +0.5,1.3 * \i + 0.5*1.3*\h/1.6 - 0.5*1.3*\i/1.6) -- (5.8 ,1.3 * \h);}
			}
		}
		
		\foreach \i in {-2,...,2} {
			\foreach \h in {0,...,1} {
				\ifthenelse{\i=0}{}{\draw[black, ->-=0.95, line width=1pt](6.2 ,1.3 * \i) -- (7.8 ,1.25 * \h+3);}
			}
		}
		\foreach \i in {-2,...,2} {
			\foreach \h in {0,...,1} {
				\ifthenelse{\i=0}{}{\draw[black, ->-=0.95, line width=1pt](6.2 ,1.3 * \i) -- (7.8 ,1.25 * \h+0.75);}
			}
		}
		\foreach \i in {-2,...,2} {
			\foreach \h in {0,...,1} {
				\ifthenelse{\i=0}{}{\draw[black, ->-=0.95, line width=1pt](6.2 ,1.3 * \i) -- (7.8 ,-1.25 * \h - 0.75);}
			}
		}
		\foreach \i in {-2,...,2} {
			\foreach \h in {0,...,1} {
				\ifthenelse{\i=0}{}{\draw[black, ->-=0.95, line width=1pt](6.2 ,1.3 * \i) -- (7.8 ,-1.25 * \h - 3);}
			}
		}
		
		\path (3.8,0) -- (4.2,0) node [black, font=\Huge, midway, sloped] {\textbf{$...$}};
		
		\foreach \i in {-2,...,3} {
    		\draw[fill=mycolor1] (0,0.8*\i-0.8*0.5) circle (0.3);  %  Input Layer_i
    	}
		
		\draw[fill=mycolor2] (2,-2.6) circle (0.3);  %  Hidden Layer1
		\path (2,-2.6) -- (2,-1.3) node [black, font=\Huge, midway, sloped] {$...$};
		\draw[fill=mycolor2] (2,-1.3) circle (0.3);  %  Hidden Layer1
		\path (2,-1.3) -- (2,0) node [black, font=\Huge, midway, sloped] {$...$};
		\draw[fill=mycolor2] (2,0) circle (0.3);  %  Hidden Layer1
		\path (2,0) -- (2,1.3) node [black, font=\Huge, midway, sloped] {$...$};
		\draw[fill=mycolor2] (2,1.3) circle (0.3);  %  Hidden Layer1
		\path (2,1.3) -- (2,2.6) node [black, font=\Huge, midway, sloped] {\textbf{$...$}};
		\draw[fill=mycolor2] (2,2.6) circle (0.3);  %  Hidden Layer1
		
		\draw[fill=mycolor2] (6,-2.6) circle (0.3);  %  Hidden Layer5
		\path (6,-2.6) -- (6,-1.3) node [black, font=\Huge, midway, sloped] {$...$};
		\draw[fill=mycolor2] (6,-1.3) circle (0.3);  %  Hidden Layer5
		\path (6,-1.3) -- (6,0) node [black, font=\Huge, midway, sloped] {$...$};
		\draw[fill=mycolor2] (6,0) circle (0.3);  %  Hidden Layer5
		\path (6,0) -- (6,1.3) node [black, font=\Huge, midway, sloped] {$...$};
		\draw[fill=mycolor2] (6,1.3) circle (0.3);  %  Hidden Layer5
		\path (6,1.3) -- (6,2.6) node [black, font=\Huge, midway, sloped] {\textbf{$...$}};
		\draw[fill=mycolor2] (6,2.6) circle (0.3);  %  Hidden Layer5
		
		\draw[fill=mycolor8] (8,-4.25) circle (0.25);  %  Output Layer
		\path (8,-4.25) -- (8,-3) node [black, font=\Huge, midway, sloped] {$...$};
		\draw[fill=mycolor8] (8,-3) circle (0.25);  %  Output Layer
		\draw[fill=mycolor9] (8,-2) circle (0.25);  %  Output Layer
		\path (8,-2) -- (8,-0.75) node [black, font=\Huge, midway, sloped] {$...$};
		\draw[fill=mycolor9] (8,-0.75) circle (0.25);  %  Output Layer
		\draw[fill=mycolor5] (8,0.75) circle (0.25);  %  Output Layer
		\path (8,0.75) -- (8,2) node [black, font=\Huge, midway, sloped] {\textbf{$...$}};
		\draw[fill=mycolor5] (8,2) circle (0.25);  %  Output Layer
		\draw[fill=mycolor6] (8,3) circle (0.25);  %  Output Layer
		\path (8,3) -- (8,4.25) node [black, font=\Huge, midway, sloped] {\textbf{$...$}};
		\draw[fill=mycolor6] (8,4.25) circle (0.25);  %  Output Layer
		
		\draw[->, line width=1pt] (-1,-0.8*2-0.8*0.5) node [font=\Large, above] {$\gamma_\mathrm{UT}$} -- (-0.3,-0.8*2-0.8*0.5);
		\draw[->, line width=1pt] (-1,-0.8*1-0.8*0.5) node [font=\Large, above] {$\beta_\mathrm{UT}$} -- (-0.3,-0.8*1-0.8*0.5);
		\draw[->, line width=1pt] (-1,-0.8*0-0.8*0.5) node [font=\Large, above] {$\alpha_\mathrm{UT}$} -- (-0.3,-0.8*0-0.8*0.5);
		\draw[->, line width=1pt] (-1,0.8*1-0.8*0.5) node [font=\Large, above] {$z_\mathrm{UT}$} -- (-0.3,0.8*1-0.8*0.5);
		\draw[->, line width=1pt] (-1,0.8*2-0.8*0.5) node [font=\Large, above] {$y_\mathrm{UT}$} -- (-0.3,0.8*2-0.8*0.5);
		\draw[->, line width=1pt] (-1,0.8*3-0.8*0.5) node [font=\Large, above] {$x_\mathrm{UT}$} -- (-0.3,0.8*3-0.8*0.5);
		
		\node[draw=white,circle,cross,minimum width=0.8 cm] at (2,1.3) {};
		\node[draw=white,circle,cross,minimum width=0.8 cm] at (6,0){};
		
		\node[align=left] at (0,-5.1) {Input \\ Layer };
		\node[align=left] at (2,-5.1) {Hidden \\ Layer 1};
		\node[align=left] at (6,-5.1) {Hidden \\ Layer $N_{h}$};
		\node[align=left] at (8,-5.1) {Output \\ Layer};
		
		\draw[->, line width=1pt, mycolor8] (8.25,-4.25) -- (9.5,-4.25) node [font=\Large, above right = -2pt and -42pt] {$o^{r^E}_{N_\mathrm{UT_V}}$};
		\draw[->, line width=1pt, mycolor8] (8.25,-3) -- (9.5,-3) node [font=\Large, above right = -2pt and -42pt] {$o^{r^E}_{1}$};
		\draw[->, line width=1pt, mycolor9] (8.25,-2) -- (9.5,-2) node [font=\Large, above right = -2pt and -42pt] {$o^{r^A}_{N_\mathrm{UT_H}}$};
		\draw[->, line width=1pt, mycolor9] (8.25,-0.75) -- (9.5,-0.75) node [font=\Large, above right = -2pt and -42pt] {$o^{r^A}_{1}$};
		
		\draw[->, line width=1pt, mycolor5] (8.25,0.75) -- (9.5,0.75) node [font=\Large, above right = -2pt and -42pt] {$o^{t^E}_{N_\mathrm{AP_V}}$};
		\draw[->, line width=1pt, mycolor5] (8.25,2) -- (9.5,2) node [font=\Large, above right = -2pt and -42pt] {$o^{t^E}_{1}$};
		\draw[->, line width=1pt, mycolor6] (8.25,3) -- (9.5,3) node [font=\Large, above right = -2pt and -42pt] {$o^{t^A}_{N_\mathrm{AP_H}}$};
		\draw[->, line width=1pt, mycolor6] (8.25,4.25) -- (9.5,4.25) node [font=\Large, above right = -2pt and -42pt] {$o^{t^A}_{1}$};
		
		\draw[->, line width=1pt, mycolor7] (10.5,-3.5) -- (11.75,-3.5) node [font=\Large, above right = -2pt and -37pt] {$o^r_{N_\mathrm{UT}}$};
		\draw[->, line width=1pt, mycolor7] (10.5,-1.5) -- (11.75,-1.5) node [font=\Large, above right = -2pt and -37pt] {$o^r_{1}$};
		\draw[->, line width=1pt, mycolor3] (10.5,1.5) -- (11.75,1.5) node  [font=\Large, above right = -2pt and -37pt] {$o^t_{N_\mathrm{AP}}$};
		\draw[->, line width=1pt, mycolor3] (10.5,3.5) -- (11.75,3.5) node [font=\Large, above right = -2pt and -37pt] {$o^t_{1}$};
		
		\path (10.8,2) -- (10.8,3.5) node [mycolor3, font=\Huge, midway, sloped] {\textbf{$...$}};
		\path (10.8,-1) -- (10.8,-3.5) node [mycolor7, font=\Huge, midway, sloped] {\textbf{$...$}};
		
		\coordinate (dm1) at (10,0.5);
        \coordinate (dm2) at (10,4.5);
        \node[rectangle, draw, mycolor3, fill=mycolor3!5, very thick, minimum width=1cm] [fit = (dm1) (dm2)] (bx4) {};
		\node[align=center,font=\normalsize,rotate=90] at (bx4.center) {Outer product + Flatten};% font=\large
		
		\coordinate (dm3) at (10,-0.5);
        \coordinate (dm4) at (10,-4.5);
        \node[rectangle, draw, mycolor7, fill=mycolor7!5, very thick, minimum width=1cm] [fit = (dm3) (dm4)] (bx44) {};
		\node[align=center,font=\normalsize,rotate=90] at (bx44.center) {Outer product + Flatten};% font=\large
		
		\coordinate (dm5) at (12.25,4.5);
        \coordinate (dm6) at (12.25,-4.5);
        \node[rectangle, draw, mycolor4, fill=mycolor4!5, very thick, minimum width=1cm] [fit = (dm5) (dm6)] (bx44) {};
		\node[align=center,font=\large,rotate=90] at (bx44.center) {Outer product + Flatten};  
		
		\draw[->, line width=1pt, mycolor4] (12.75,-2.95) -- (13.75,-2.95) node [font=\Large, above right = -2pt and -17pt] {$o_{N_\mathrm{AP}, N_\mathrm{UT}}$};
		\draw[->, line width=1pt, mycolor4] (12.75,-1.65) -- (13.75,-1.65) node [font=\Large, above right = -2pt and -17pt] {$o_{N_\mathrm{AP}, 1}$};
		\draw[->, line width=1pt, mycolor4] (12.75,-0.35) -- (13.75,-0.35) node [font=\Large, above right = -2pt and -17pt] {$o_{2, N_\mathrm{UT}}$};
		\draw[->, line width=1pt, mycolor4] (12.75,0.95) -- (13.75,0.95) node [font=\Large, above right = -2pt and -17pt] {$o_{2, 1}$};
		\draw[->, line width=1pt, mycolor4] (12.75,1.65) -- (13.75,1.65) node [font=\Large, above right = -2pt and -17pt] {$o_{1, N_\mathrm{UT}}$};
		\draw[->, line width=1pt, mycolor4] (12.75,2.95) -- (13.75,2.95) node [font=\Large, above right = -2pt and -17pt] {$o_{1, 1}$};
		
		\path (13.05,1.65) -- (13.05,2.95) node [mycolor4, font=\Huge, midway, sloped] {\textbf{$...$}};
		\path (13.05,-0.35) -- (13.05,0.95) node [mycolor4, font=\Huge, midway, sloped] {\textbf{$...$}};
		\path (13.05,-1.65) -- (13.05,-0.35) node [mycolor4, font=\Huge, midway, sloped] {\textbf{$...$}};
		\path (13.05,-2.95) -- (13.05,-1.65) node [mycolor4, font=\Huge, midway, sloped] {\textbf{$...$}};
  		\end{tikzpicture}}
	\caption{The extended multi-task structured deep neural network with $N_\mathrm{AP_H} + N_\mathrm{AP_V} + N_\mathrm{UT_H} + N_\mathrm{UT_V}$ neurons at its output layer.}
	\label{DeepLearn-MultiTask-2}
\end{figure}

\subsection{Baselines}\label{SubSec:Baselines}
We consider the inverse fingerprinting (IFP) beam selection method as a baseline to compare the proposed deep learning-based methods with probabilistic approaches \cite{va_inverse_2018}. The IFP also is a data-driven method that uses training samples to make a lookup table of best beam pairs for different environment locations. However, the IFP method in \cite{va_inverse_2018} uses only the position information of the UT, not considering the arbitrary orientation of the UT. Since the DFT-based codebook has unequal beamwidths, it is not feasible to propose a unique candidate beam list at an orientation angle $\boldsymbol{A}_{r_1}$ based on a candidate beam list obtained for a different orientation angle $\boldsymbol{A}_{r_2}$. Thus, we introduce an extended version of the IFP method \cite{va_inverse_2018}, which takes into account the orientation of the UT array in the beam selection procedure. This is a straightforward extension of the GIFP algorithm we proposed in \cite{rezaie_location-_2020}.

\subsubsection{Generalized Inverse fingerprinting Method}\label{SubSubSec:GIFP}
%As illustrated in Fig.~\ref{Generalized}, 
The bin definition of the IFP method is extended in a way that considers the rotation of the UT array about $x$,$y$, and $z$ axes. By extending discretization to both the spatial and angular domains,  position and orientation of the UT, $\boldsymbol{p}_\mathrm{UT}$ and $\boldsymbol{\psi}_\mathrm{UT}$ respectively, determine the corresponding bin for each observation. The $k{\text{-th}}$ bin may be defined as
\begin{equation}
\begin{split}
\mathcal{B}_k = & [x_k, x_k + \Delta_s) \times [y_k, y_k + \Delta_s) \times [z_k, z_k + \Delta_s) \times \\ & [\alpha_k, \alpha_k + \Delta_a) \times [\beta_k, \beta_k + \Delta_a) \times [\gamma_k, \gamma_k + \Delta_a)\\
\end{split}
\end{equation}
where $\Delta_s$ and $\Delta_a$, as hyper parameters of the generalized inverse fingerprinting (GIFP) method, are spatial bin size (SBS) and angular bin size (ABS), respectively. The values $x_k$, $y_k$, $z_k$, $\alpha_k$, $\beta_k$, and $\gamma_k$ are selected to obtain a set of bins that have no overlap and cover all the spatial and angular coordinates of interest. Note that the bin definition of the IFP method in \cite{va_inverse_2018} can be recovered by setting $\Delta_a = 2\pi$.

Following the IFP method approach in proposing a candidate list for the $k{\text{th}}$ bin, $\mathcal{S}_k$, the beam pairs of this list are chosen to minimize the misalignment probability, i.e., the probability of not including the beam pair with the highest RSS in the list. The misalignment probability may be expressed as
\begin{equation} \label{eq:Mis}
P_{m}(\mathcal{S}_k) = \mathbb{P} \left[\max\limits_{(i, j) \in \mathcal{S}_k} R_{i,j} < \max\limits_{(p, q) \in \mathcal{B}} R_{p,q}\right],
\end{equation}
where $\mathcal{B}$ is a set including all possible beam pairs. Using the training samples assigned to the $k{\text{th}}$ bin, the members of $\mathcal{S}_k$ are the top $N_b$ ranked beam pairs which have the highest frequencies of being best \cite{va_inverse_2018}. In test mode, for a given new UT coordinate and orientation, the candidate list of the associated bin is used, and the beam selection follows the procedure in Fig. \ref{fig:Sensing}\subref{fig:Sensing_CABS}. Because of the discretization in the angular domain added to the spatial domain, there are significantly more bins in GIFP compared to classical IFP. Thus, the GIFP method becomes more data-demanding, and it needs a much larger training dataset to have the same density of training samples over bins as IFP.

\subsubsection{Hierarchical Beam Search}\label{SubSubSec:HBS}
\begin{figure}[t]
	\centering
	\subfloat[$k_h = 0, k_v = 1$\label{1a}]{%
       \scalebox{1}{\includegraphics[width=0.45\textwidth]{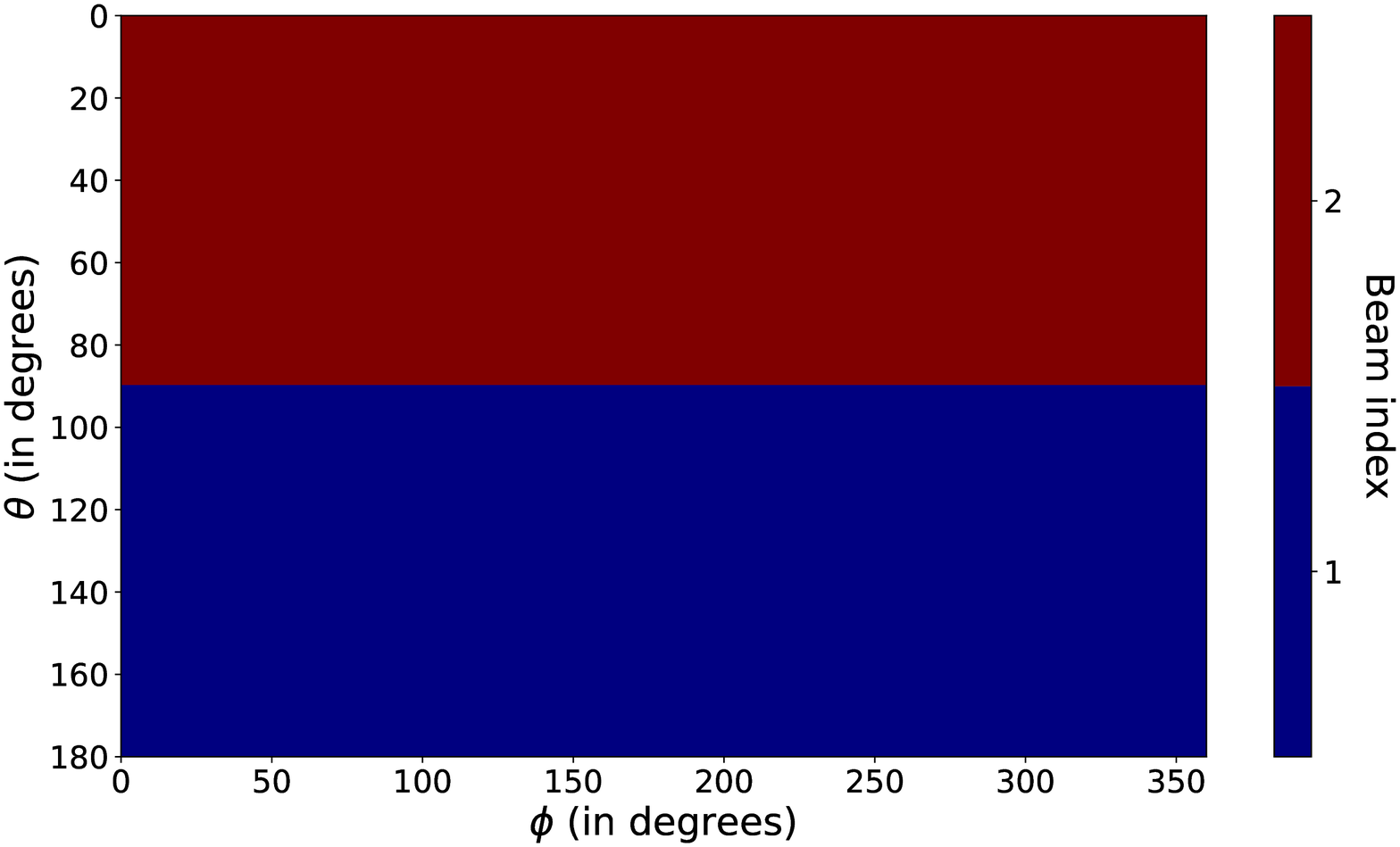}}}
    \hspace{2em}%
  \subfloat[$k_h = 0, k_v = 2$\label{1b}]{%
        \scalebox{1}{\includegraphics[width=0.45\textwidth]{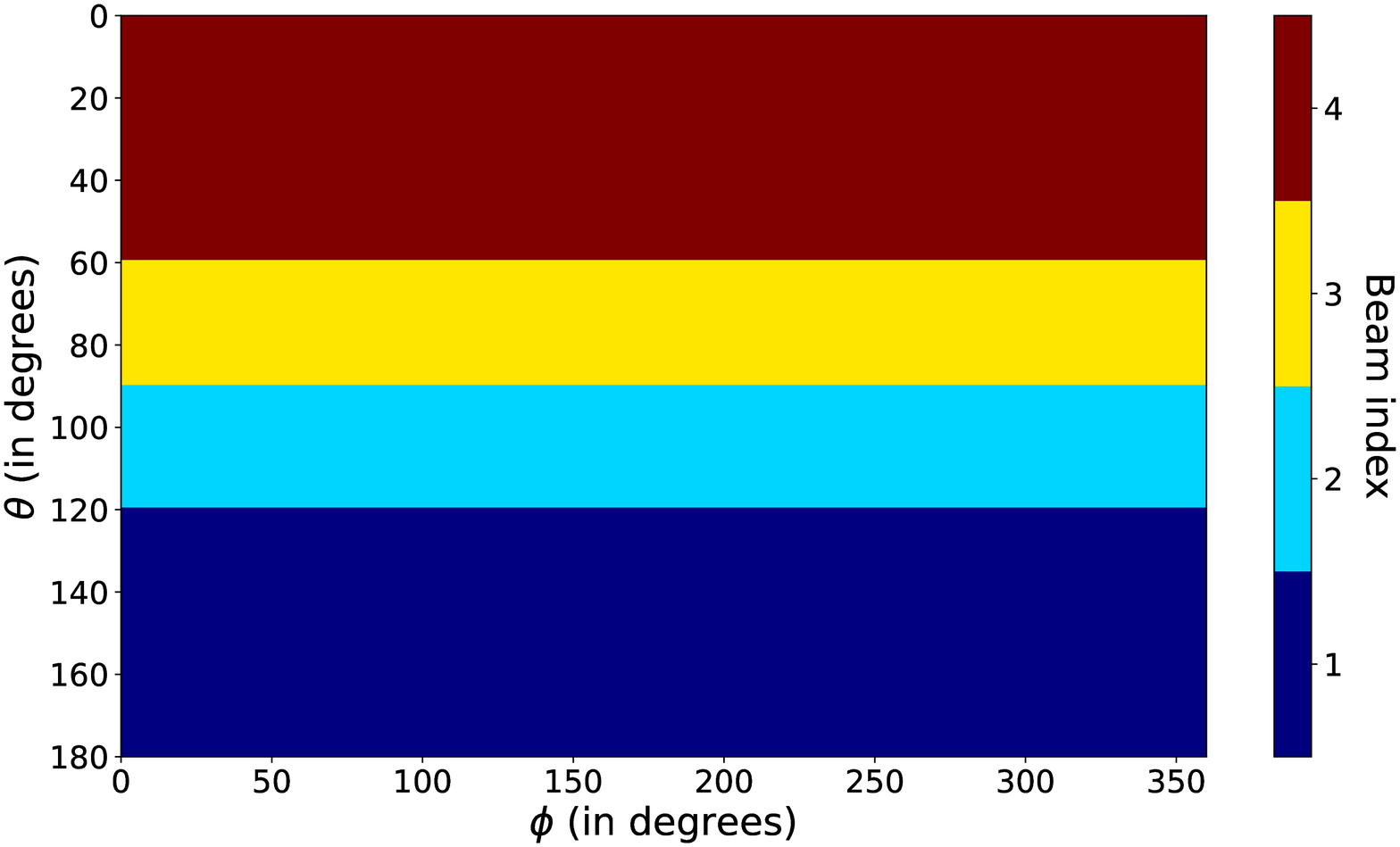}}}
    \\
  \subfloat[$k_h = 1, k_v = 2$\label{1c}]{%
        \scalebox{1}{\includegraphics[width=0.45\textwidth]{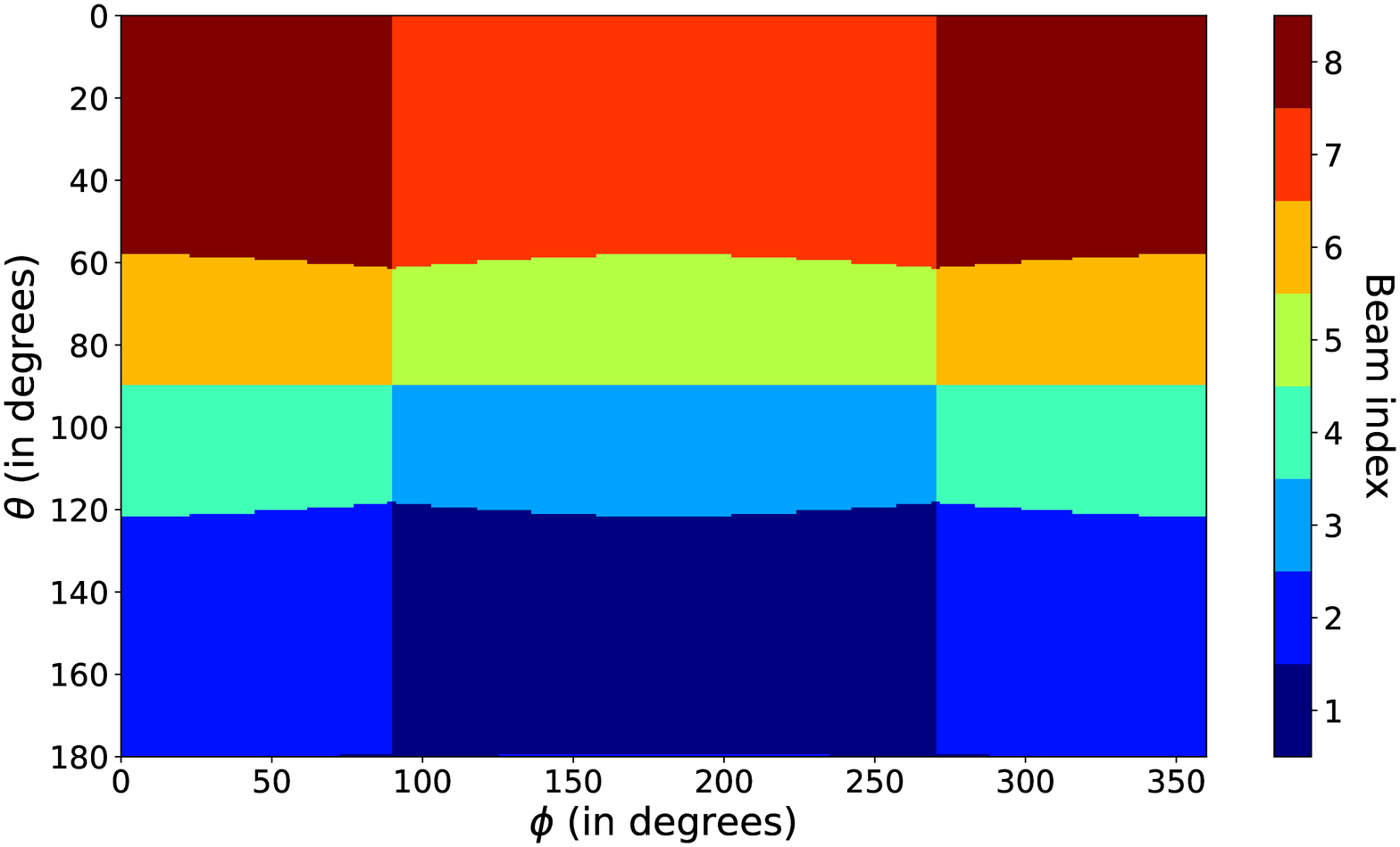}}}
    \hspace{2em}%
  \subfloat[$k_h = 2, k_v = 2$\label{1d}]{%
        \scalebox{1}{\includegraphics[width=0.45\textwidth]{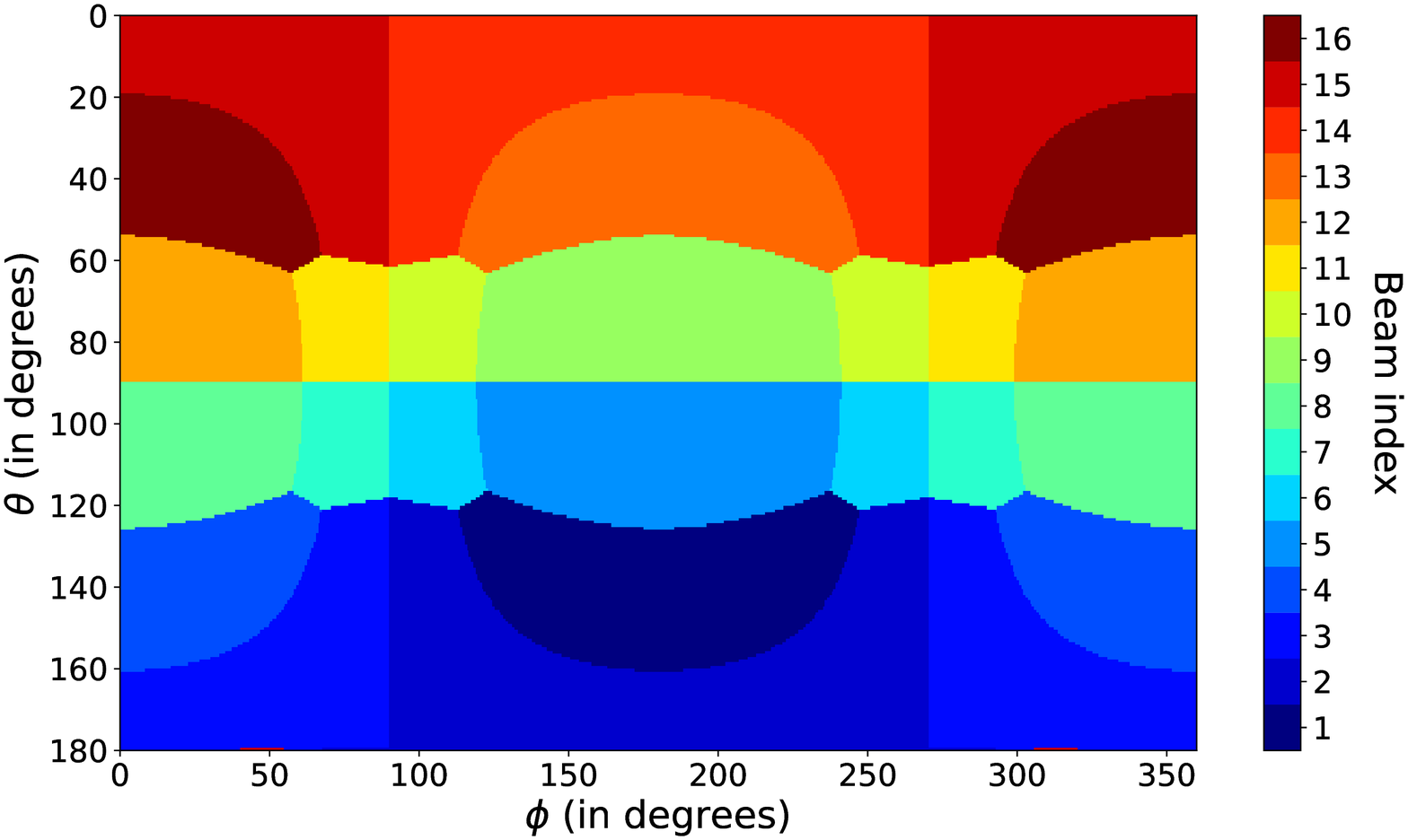}}}
	\caption{Beam decision regions in different levels of the DEACT method as a hierarchical beam search. The antenna array size is $\{N_{h} = 4, N_{v}=4\}$. At each level of the search, different colors are representing different codewords in the codebook.}
	\label{Fig:HBS}
\end{figure}

As mentioned before, in hierarchical beam search, the space is covered and sensed in several stages with different beam width. For a UPA with $N = N_h * N_v$ antennas, there are $\log_2(N_h) + 1$ and $\log_2(N_v) + 1$ levels in horizontal and vertical directions. At the $k_h{\text{th}}$ and $k_v{\text{th}}$ horizontal and vertical level, there are $N_{k_h} \times N_{k_v}$ codewords, where $N_{k_h} = 2^{k_h}$ and $ N_{k_v} = 2^{k_v}$. By extending the DEACT approach proposed in \cite{xiao_hierarchical_2016}, the $n_h{\text{th}}$ and $n_v{\text{th}}$ codeword in the $k_h{\text{th}}$ and $k_v{\text{th}}$ horizontal and vertical level of the codebook can be written as
\begin{equation}
\begin{split}
    \boldsymbol{w}(k_h, k_v, n_h, n_v)& = [\boldsymbol{a}^T_\mathrm{E}(N_{k_v}, \theta_v), \boldsymbol{0}^T_{(N_v - N_{k_v})\times 1}]^T \otimes [\boldsymbol{a}^T_\mathrm{A}(N_{k_h}, \phi_h, \theta_v), \boldsymbol{0}^T_{(N_h - N_{k_h})\times 1}]^T, \\
    &\phi_h = \arccos\frac{2n_h-1-N_{k_h}}{N_{k_h}}, \hspace{4pt} \theta_v = \arccos\frac{2n_v-1-N_{k_v}}{N_{k_v}},
\end{split}
\end{equation}
where $\boldsymbol{a}_\mathrm{A}$ and $\boldsymbol{a}_\mathrm{E}$ are defined in \eqref{eq:a_A} and \eqref{eq:a_E}. Algorithm~\ref{alg:VH_Beamforming} illustrates the procedure of horizontal and vertical beamforming using a 3D codebook. The device first searches the vertical codewords while considering omnidirectional horizontal beamforming during this part. Then, the device tries to find the best horizontal codeword with the knowledge of the best vertical one. At each stage of the HBS, the beam decision region of a codeword is the subspace in the angular domain where the specified codeword has the highest gain among all the possible codewords. As an example, Fig. \ref{Fig:HBS} shows the beam decision regions for different levels of the DEACT method for a UPA with $N_h = 4$ and $N_v = 4$.

\RestyleAlgo{ruled}
\begin{algorithm}
    \caption{Vertical and Horizontal Analog beamforming Based on Codebook}\label{alg:VH_Beamforming}
    \textbf{Input:}  $N_h$, $N_v$, $\boldsymbol{w}$\\
    \KwResult{$n_h, n_v$}
    \textbf{Initialize:}  Omnidirectional horizontal beamforming ($k_h = 0$), $n_h = 1$, $n_v = 1$\\
    \For{$k_v = 1$ to $\log_2(N_v) + 1$}{
     %1. Select the 2 candidate codewords with indices $n^{(1)}_{v}$ and $n^{(2)}_{v}$ among the possible option at level $k_v$ and results in previous levels
     1. $n^{(1)}_{v} = 2 n_v -1$ , $n^{(2)}_{v} = 2 n_v$\\
     2. Sense the environment with the two selected codewords:  $\boldsymbol{w}(k_h, k_v, n_h, n^{(1)}_{v})$ and $\boldsymbol{w}(k_h, k_v, n_h, n^{(2)}_{v})$\\
     %3. Select the codeword with the higher RSS\\
     \eIf{$\boldsymbol{w}(k_h, k_v, n_h, n^{(1)}_{v})$ provides higher RSS}{
      $n_v = n^{(1)}_{v}$
      }{$n_v = n^{(2)}_{v}$}
     }
     \For{$k_h = 1$ to $\log_2(N_h) + 1$}{
     1. $n^{(1)}_{h} = 2 n_h - 1$ , $n^{(2)}_{h} = 2 n_h$\\
     2. Sense the environment with the two selected codewords:  $\boldsymbol{w}(k_h, k_v, n^{(1)}_{h}, n_v)$ and $\boldsymbol{w}(k_h, k_v, n^{(2)}_{h}, n_v)$\\
     %3. Select the codeword with the higher RSS\\
     \eIf{$\boldsymbol{w}(k_h, k_v, n^{(1)}_{h}, n_v)$ provides higher RSS}{
      $n_h = n^{(1)}_{h}$
      }{$n_h = n^{(2)}_{h}$}
     }
\end{algorithm}

\section{Simulation Results and Discussions}\label{Sec:SimResults}
In this section, we evaluate the performance of different beam alignment methods in different situations. We start by introducing the ray-tracing setup. Following we compare the performance of the methods in terms of misalignment probability and effective spectral efficiency (ESE). In order to quantify the training data requirements of each of the analyzed data-driven methods, we investigate their performance when they are trained using datasets of different sizes. Finally, we evaluate the robustness of the deep learning based and GIFP methods with inaccurate position and orientation information.

\subsection{Simulation Setup and Performance Measures}
\begin{figure}[t]
	\centering
	\scalebox{0.15}{
		\includegraphics[trim={0cm 0cm 0cm 0cm},clip]{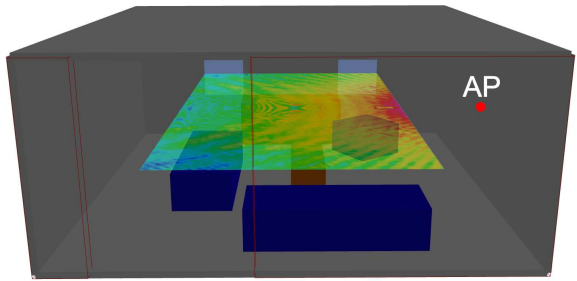}
	}
	\caption{The ray-tracing simulation of the living room as the simulation scenario. The AP position is shown with the red circle. Also, the received LOS power is depicted in the user grid of the scenario.}
	\label{Fig:LR}
\end{figure}

We consider the living room (LR) described and defined precisely in the IEEE 802.11ad task group \cite{maltsev_channel_nodate}. Fig. \ref{Fig:LR} illustrates the LR as the considered indoor environment. According to the standard, the LR dimensions are $7$m $\times 7$m $\times 3$m (W$\times$ L$\times$ H). Two sofas, a table, and an armchair are placed in the LR, and a cabinet is placed between two windows of one of the outer walls. The AP is placed in the middle of one of the LR walls. The UT can take a position in a sector with the same height as the AP, where the sector has dimensions of $4$m $\times 7$m (W$\times$ L). In the ray tracing tool, we defined $70,000$ UT positions in the user sector. In the standard, a cluster blockage model is considered where a NLOS path can be blocked with a probability that is defined in the standard. This probability depends on the order of reflection and the reflective surfaces. In addition, the probability of LOS blockage is considered a model parameter that can be set to $0$ or $1$. To construct datasets with arbitrary LOS blockage probability, we can combine the samples proportionally from datasets with $0$ and $1$ LOS blockage probability. Table \ref{T1} gives a summary of the most important information about the LR. 

\begin{table}
    \caption{Geometrical Information of the Living Room}
    \label{T1}
    \centering
    \scalebox{0.8}{
    \begin{tabular}{ c  c  c  c  c}
        \hline \hline
        & Dimension (m) & AP's pos. (m) & User grid (m) & $\#$ UT positions\\
        \hline
        LR & $7\times7\times3$ & $(7, 3.5, 1.5)$ & $[1.5-5.5] \times [0-7] \times [1.5]$ & 70,000\\
        \hline
    \end{tabular}}
\end{table}

To generate channel responses, we use Altair Feko-Winprop software \cite{noauthor_altair_nodate} as a professional ray-tracing software, which considers empirical losses for transmission, reflection and diffraction in the simulations. Using the outputs of the ray-tracing tool including the AoD, AoA, and path gains of all the paths, the channel response can be constructed using \eqref{Eq:Channel}. The $25$ strongest multipath components are used at any UT position to generate the channel response.

For all samples of the dataset, we consider $P_\mathrm{AP}=0\hspace{2pt}$dBm and $\sigma_n^2=-84\hspace{2pt}$dBm as the transmit power and noise variance, respectively. We consider $\{8, 8\}$ and $\{4, 4\}$ UPA at the AP and UT, respectively. After constructing the channel response for each UT location, we calculate the RSS for each beam pair using \eqref{Eq:RSS}. According to the type of encoding scheme, $M$, the labels are calculated and stored in $\mathbb{D}^{\Xi}$. We consider $80\%$ of each dataset for the training scheme and the remaining $20\%$ as the test samples. To reduce the dependencies between the training samples and the performance of the data-driven methods, we use the $5-$fold technique. %In this technique, each experiment is repeated $5$ times with a different training and test samples selection. Indeed, the dataset is split into $5$ parts, where at each repetition of the experiment, one of the parts is selected as the test data and the rest as the training data.
In this technique, the dataset is divided into $5$ parts. Then, the training and testing experiments are repeated $5$ times, each of them with a different part of the datset selected as test data, with the end result obtained by averaging over all $5$ experiments.

There are $70,000$ UT positions in the user grid of the LR, which are stored in $\mathbb{D}^{\Xi}$. To evaluate the performance of the data-driven methods with limited training datasets, we make two sub-datasets $\mathbb{D}^{\Xi}_{10\%}$ and $\mathbb{D}^{\Xi}_{1\%}$, respectively, consisting of $7,000$ and $700$ samples randomly selected from $\mathbb{D}^{\Xi}$. In the performance evaluations, we consider $5$ beam alignment approaches:
\begin{enumerate}
    \item \textbf{DNN-ST}: The DNN method with single-task structure described in Section~\ref{SubSubSec:ST},
    \item \textbf{DNN-MT}: The DNN method with multi-task structure explained in Section~\ref{SubSubSec:MT},
    \item \textbf{DNN-EMT}: The DNN method with extended multi-task structure described in Section~\ref{SubSubSec:EMT},
    \item \textbf{GIFP}: The generalized inverse fingerprinting method explained in Section~\ref{SubSubSec:GIFP},
    \item \textbf{HBS}: The hierarchical beam search method with the deactivation technique represented in Section~\ref{SubSubSec:HBS}.
\end{enumerate}
All the neural networks structures have $N_h = 5$ hidden layers with $n_h = 128$ neurons at each hidden layer. Also, to prevent overfitting in the neural networks, $10\%$ dropout for all the hidden layers is used. We use Adam optimizer \cite{kingma_adam_2015} in the training phase with $50$ epochs, while the minibatch size is progressively increased from $32$ to $8192$ samples \cite{rezaie_location-_2020}. To reduce the effects of initial weights of neural networks on the beam alignment performance, we averaged over $3$ results with $3$ random weight initializations for each experiment. To reproduce the numerical results, the code and datasets are released at \href{https://github.com/SajadRezaie/3DOriLocBeamSelection.git}{\url{https://github.com/SajadRezaie/3DOriLocBeamSelection}}.

\begin{figure}[t]
\centering
\scalebox{0.7}{
\begin{tikzpicture}
\coordinate (dm0) at (8,2);
\coordinate (dm1) at (0,0);
\coordinate (dm2) at (0.5,0);
\coordinate (dm3) at (1,0);
\coordinate (dm33) at (2,2);
\coordinate (dm4) at (2,0);
\coordinate (dm5) at (2.5,0);

\draw[draw=white!40!black, fill = cyan] (dm1) rectangle ++(0.5,2) node[pos=.5, rotate=90] {$b_1$};
\draw[draw=white!40!black, fill = cyan] (dm2) rectangle ++(0.5,2) node[pos=.5, rotate=90] {$b_2$};
\draw[draw=white!40!black, fill = cyan] (dm3) rectangle (dm33) node[pos=.5, rotate=0] {$\cdots$};
\draw[draw=white!40!black, fill = cyan] (dm4) rectangle ++(0.5,2) node[pos=.5, rotate=90] {$b_{N_b}$};
\draw[draw=white!40!black, fill = magenta] (dm5) rectangle (dm0) node[pos=.5, rotate=0] {Data};

%\draw[thick, draw=black] (dm1) rectangle (dm0);
\draw[line width=0.5mm,->] (-0.1,-0.1) -- ++(9,0) node [right] {Time};
\draw[line width=0.25mm,<->] (0,2.1) -- (8,2.1)node[pos=.5, rotate=0, above] {$T_{fr}$};
\draw[line width=0.25mm,<->] (0,2.3) -- (0.5,2.3)node[pos=.5, rotate=0, above] {$T_{s}$};

\end{tikzpicture}}
\caption{Frame structure of the beam alignment procedures. In the beam selection method, the transceivers sense the environment with $N_b$ beam pairs, and then use the rest of the frame to transmit the actual data.}
\label{Fig:DataFrame}
\end{figure}
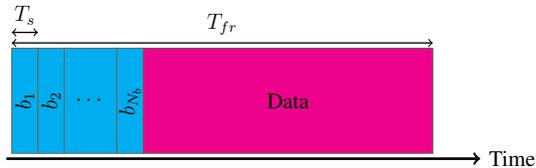

For a channel with a given coherence time, the channel resources used for the beam alignment phase should be deducted from the achievable data rate of the system \cite{hussain_second-best_2019}. 
Fig.~\ref{Fig:DataFrame} shows the frame structure of the initial beam alignment procedures. $T_{fr}$,  $T_s$, and $N_b$ are the frame duration, beam scanning time of one beam pair, and the number of sensed beam pairs in the beam alignment phase, respectively. We assume the frame duration is less than the coherence time of the channel, and the channel response is not varying during the whole frame. In contrast to the beam selection methods, in the HBS approach the number of sensed beam pairs is fixed and entirely determined by the antenna array configurations as $N_b = 2\log_2(N_\mathrm{UT}) + 2\log_2(N_\mathrm{AP})$. The transceivers use the selected beam pair in the initial beam alignment phase to communicate the actual data.

In the beam selection methods, the transceivers sense the environment with all the beam pairs in the candidate list $\mathcal{S}$ ($ | \mathcal{S} | = N_b |$) and select the one that provides the highest RSS, i.e.
\begin{equation} \label{eq:i_j_max}
i^*, j^* = \underset{i, j}{\mathrm{arg\hspace{2pt}max}} \hspace{2pt} (R_{i, j}).
\end{equation}
As a performance measure, we define the misalignment probability by reformulating \eqref{eq:Mis} as:
\begin{equation} \label{eq:MisPMetric}
P_{m}(\mathcal{S}_k) = \mathbb{P} \left[R_{i^*,j^*} < \max\limits_{(p, q) \in \mathcal{B}} R_{p,q}\right]
\end{equation}
where $\mathcal{B}$ ($| \mathcal{B} | = N_\mathrm{AP} N_\mathrm{UT}$) includes all the possible beam pair combinations. For the HBS approach, $i^*, j^*$ are found by sensing the environment with different beam width, and then we use \eqref{eq:MisPMetric} to calculate the misalignment probability. In addition, the effective spectral efficiency (ESE) as another performance measure may be written as 
\begin{equation}
    \text{SE}_{\text{eff}} = \frac{T_{fr} - N_b T_s}{T_{fr}} \log_2 (1+ \text{SNR}_{i^*, j^*}), \hspace{10pt} N_b T_s \leq T_{fr}
\end{equation}
where $\text{SNR}_{i, j}$ is the SNR of $(i,j)$th beam pair, i.e.
\begin{equation}
    \text{SNR}_{i, j} = \frac{\Big \lVert \sqrt{P_\mathrm{AP}} \boldsymbol{v}_{j}^H \boldsymbol{H} \boldsymbol{u}_{i} s \Big \rVert ^2}{\sigma_n^2}.
\end{equation}
We use $T_{fr}=20$ms and $T_{s}=0.1$ms in the numerical evaluations \cite{hussain_second-best_2019}. In addition to the described baselines in Section \ref{SubSec:Baselines}, we use perfect alignment as a genie-aided method that always selects the beam pair that has the highest SNR. The perfect alignment provides an upper bound for ESE.

\subsection{Numerical Evaluation}
\begin{figure}[t]
	\centering
	\subfloat[$\mathbb{D}^{\Xi}$ with  $70,000$ samples\label{Fig:_SEDE}]{%
       \scalebox{0.899}{
       \includegraphics[width=0.45\textwidth, trim={0.68cm 0.6cm 2.3cm 1.5cm},clip]{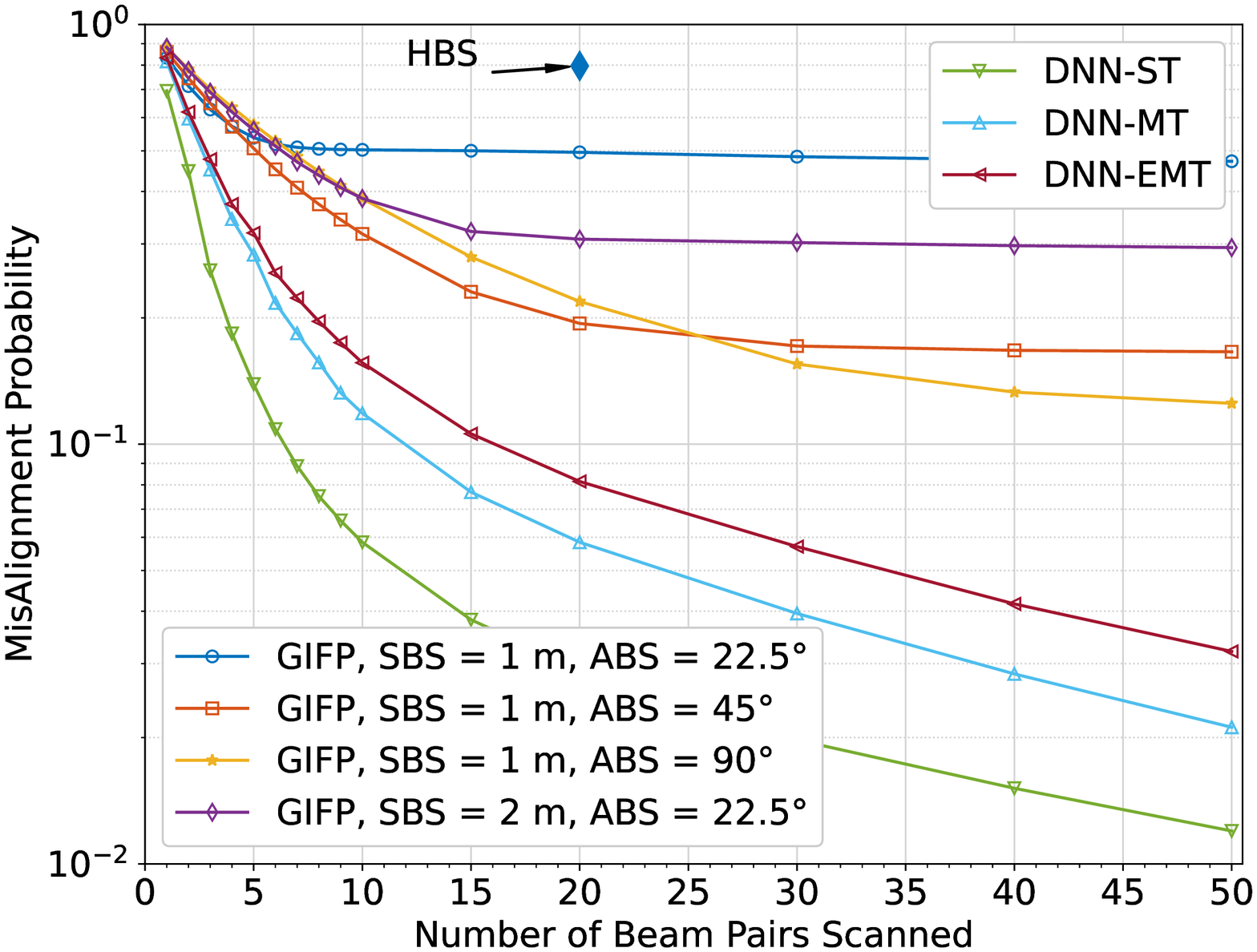}
       \hspace{2em}%
       \includegraphics[width=0.45\textwidth, trim={0.8cm 0.6cm 2.3cm 1.5cm},clip]{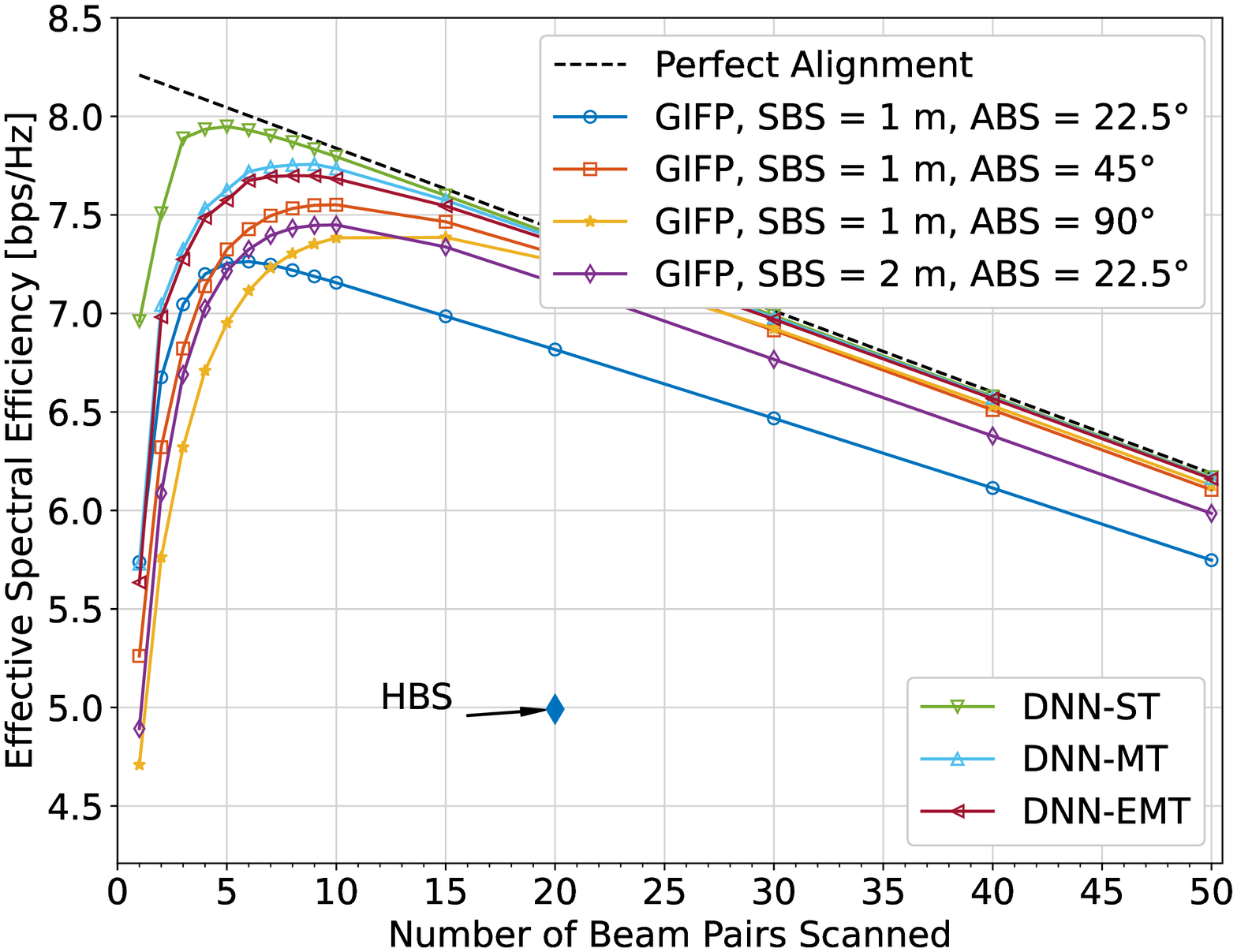}}}
    \\
    \subfloat[$\mathbb{D}^{\Xi}_{10\%}$ with $7,000$ samples\label{Fig:_SEDE10}]{%
       \scalebox{0.899}{
       \includegraphics[width=0.45\textwidth, trim={0.68cm 0.6cm 2.3cm 1.5cm},clip]{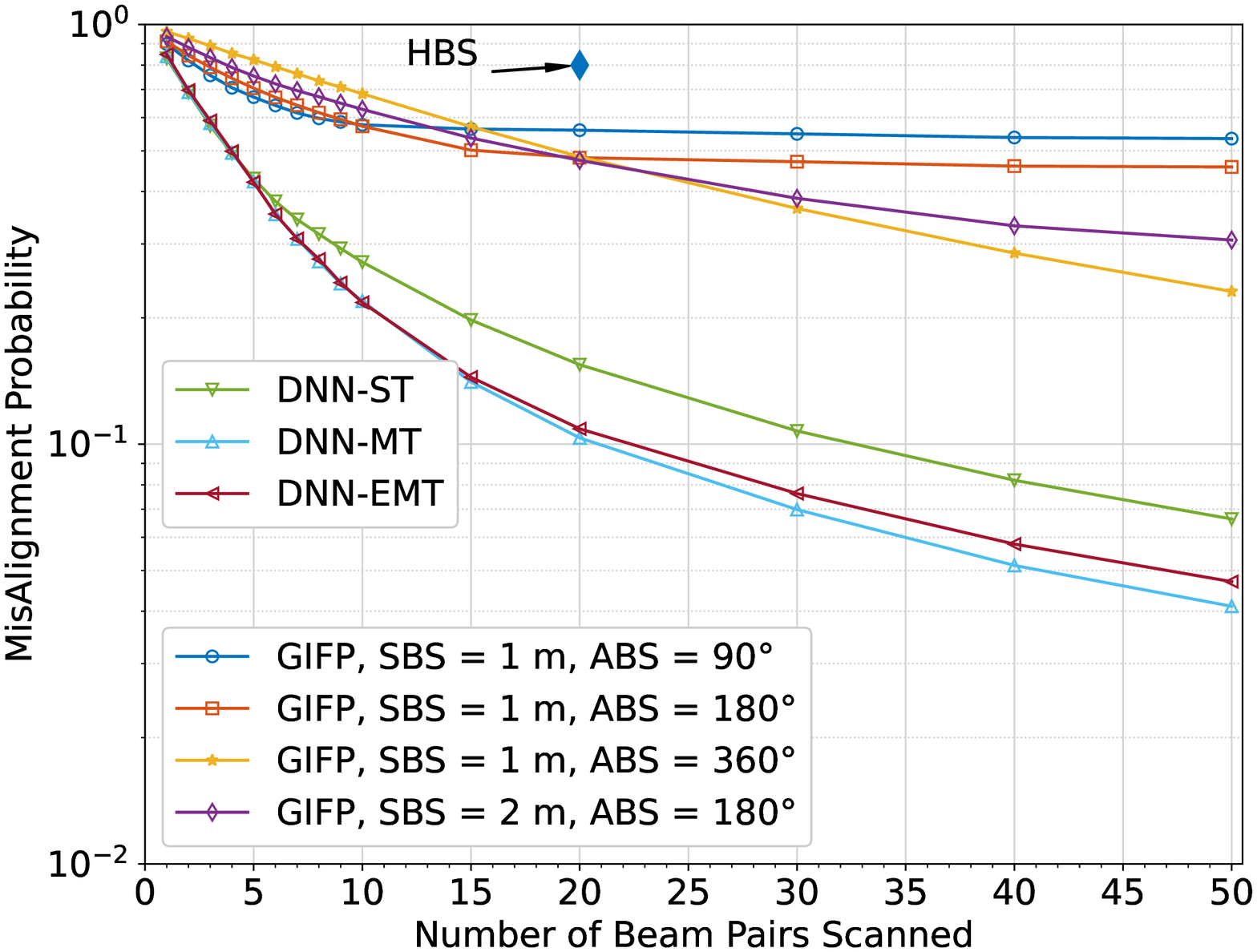}
       \hspace{2em}%
       \includegraphics[width=0.45\textwidth, trim={0.8cm 0.6cm 2.3cm 1.5cm},clip]{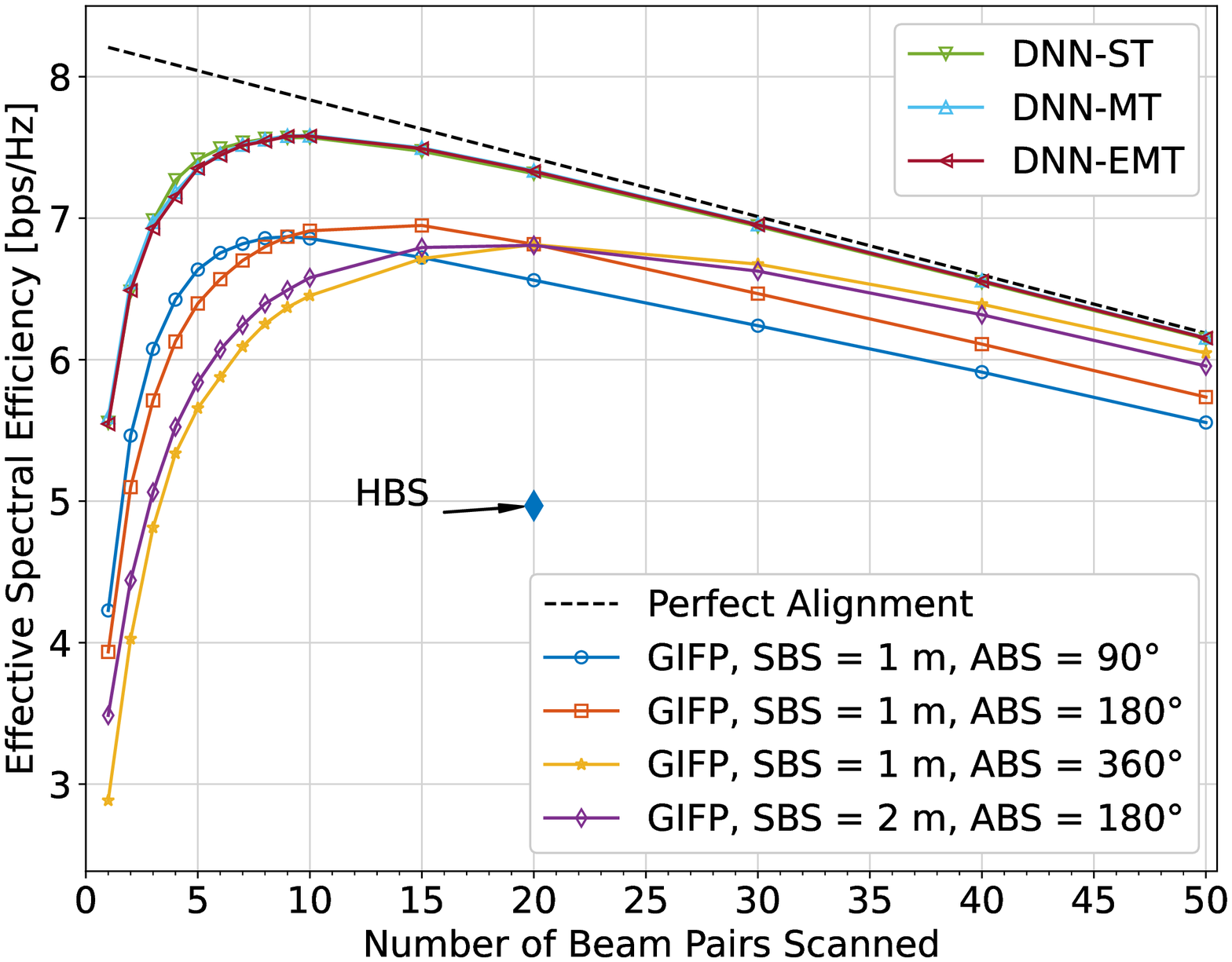}}}
    \\
    \subfloat[$\mathbb{D}^{\Xi}_{1\%}$ with $700$ samples\label{Fig:_SEDE1}]{%
       \scalebox{0.899}{
       \includegraphics[width=0.45\textwidth, trim={0.68cm 0.6cm 2.3cm 1.5cm},clip]{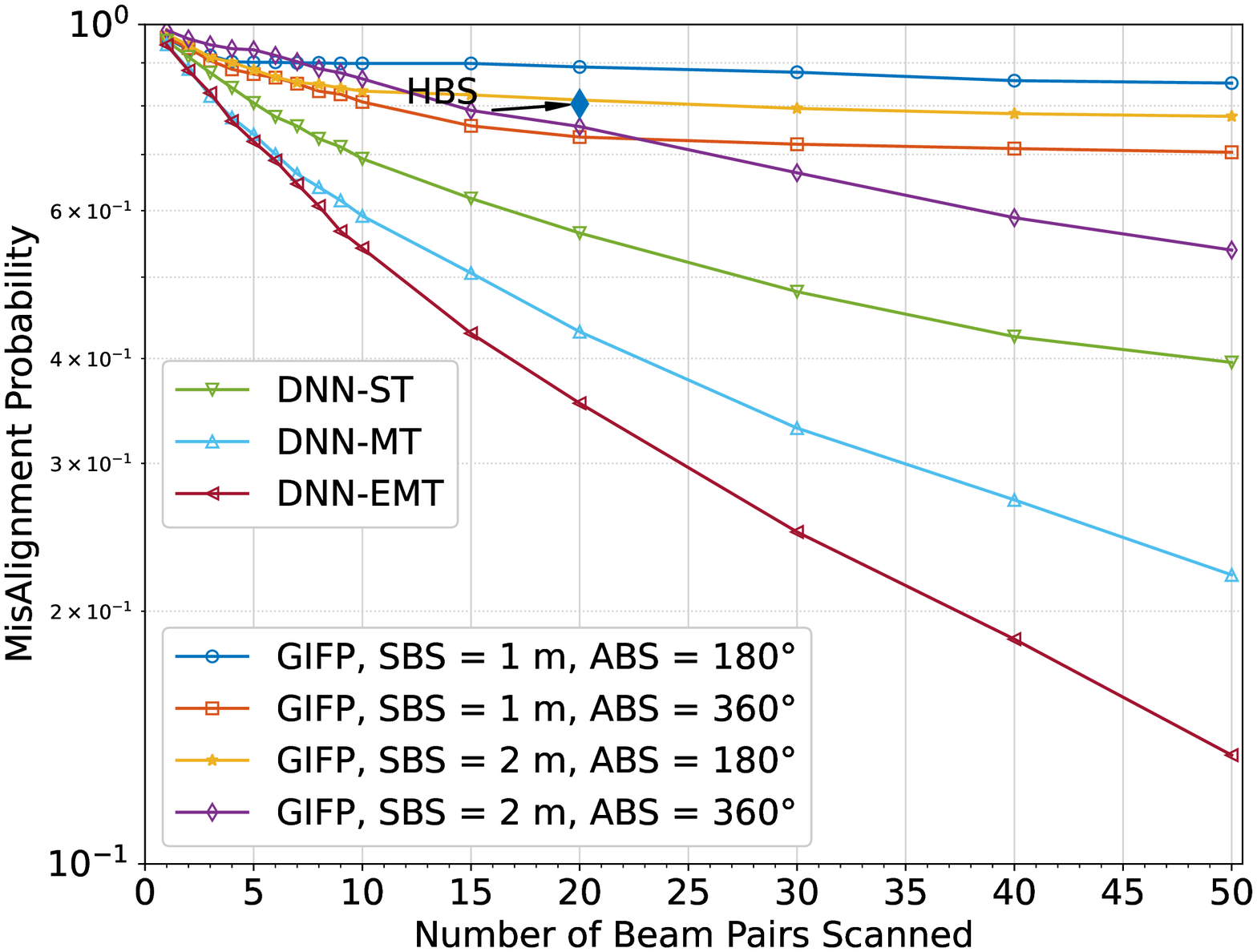}
       \hspace{2em}%
       \includegraphics[width=0.45\textwidth, trim={0.8cm 0.6cm 2.3cm 1.5cm},clip]{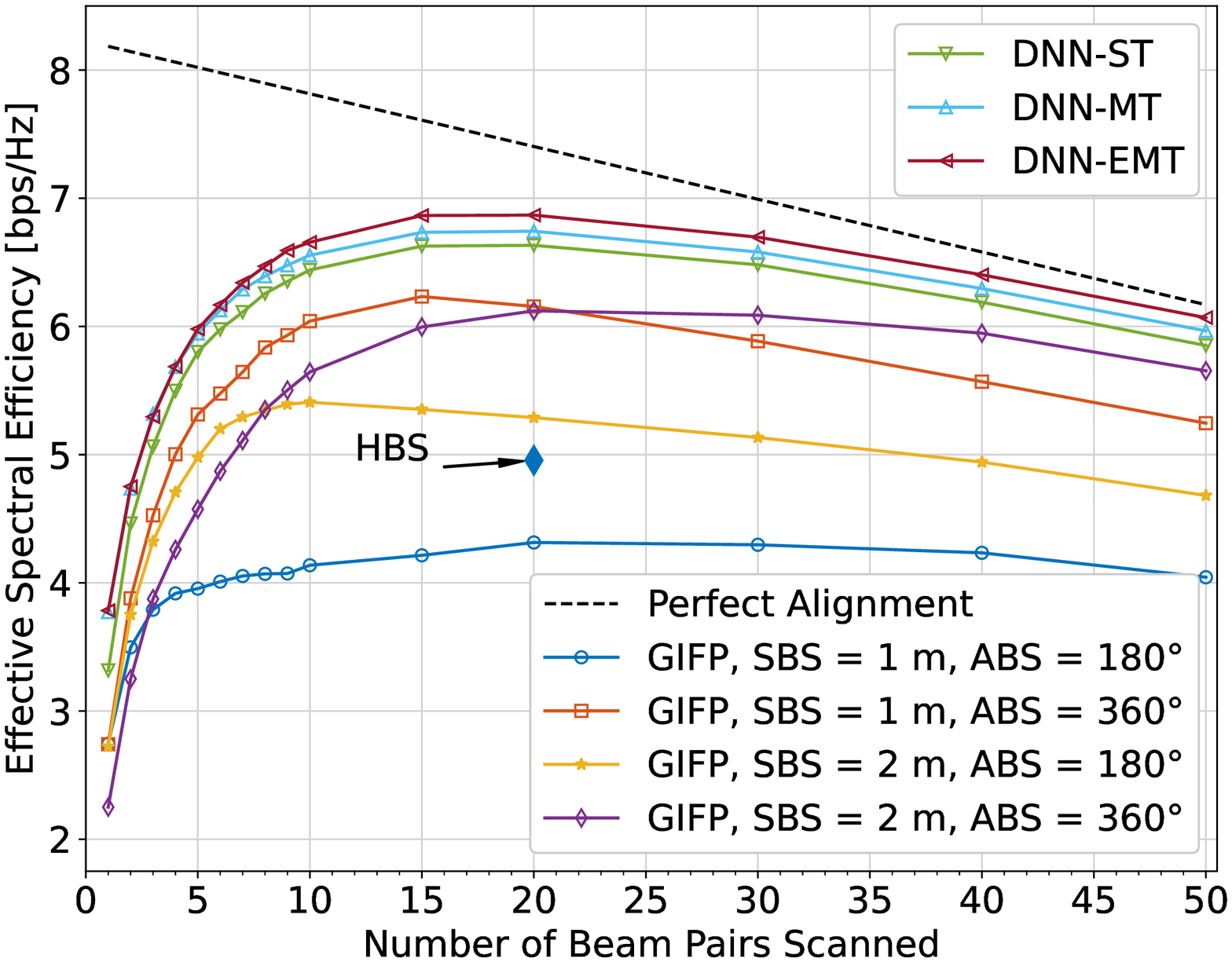}}}
	\caption{Misalignment probability and effective spectral efficiency for different dataset sizes.}
	\label{Fig:DiffDatasets}
\end{figure}

In Fig.~\ref{Fig:DiffDatasets}, the performance of the different investigated beam selection methods is depicted. The plots show the misalignment probability and spectral efficiency of the methods as a function of the the number of beam pairs scanned. For data-driven methods, this corresponds to the size $N_b$ of the candidate beam list $\mathcal{S}$. For HBS, on the other hand, the number of scanned beams is fixed and determined by the size of the hierarchical codebooks defined at AP and UT; in the selected configurations, this corresponds to $2\log_2(16) + 2\log_2(64) = 20$ beam pairs. As it can be seen from Fig. \ref{Fig:DiffDatasets}\subref{Fig:_SEDE}, the data-driven beam selection methods have much lower misalignment probability and up to $60\%$ higher ESE than the HBS method. In addition, in the data-driven beam selection methods, the candidate list may include a different number of beam pairs for scanning, $N_b$. For example, the DNN-ST structure with a candidate list size of $N_b = 3$ can reach more than $50\%$ higher ESE and $85\%$ less latency than the HBS method, which scans $20$ beam pairs. Also, all the deep-learning based methods perform significantly better than GIFP methods with different spatial and angular bin sizes. For example, the GIFP with SBS = $1$m and ABS = $22.5^{\circ}$ has $4\times7\times1\times16\times4\times4 = 7168$ bins, which end up with around $8$ samples on average at each bin. The excellent performance of DNN-ST is a consequence of training it with a large dataset of sufficient size. It clearly outperforms the DNN-MT and DNN-EMT methods, which suffer from the decoupling of the beam selection process at AP and UT for both cases, and between horizontal and vertical beam directions in the latter case. On the other hand, the DNN-MT and DNN-EMT networks have much less computational cost than the DNN-ST structure in the training and test. The computational complexity is proportional to the number of trainable parameters, and there are significantly fewer trainable weights in the DNN-MT and DNN-EMT.

As the dataset size is decreased, the performance advantages of DNN-ST progressively fade away. This is illustrated in Fig. \ref{Fig:DiffDatasets}\subref{Fig:_SEDE10} where, with a dataset size of $7000$ samples, the DNN-MT method with fewer trainable parameters has the best performance. In spite of the $90\%$ reduction of the training dataset size, the deep learning-based methods work significantly better than the HBS method. The performance of GIFP is further deteriorated with the decrease in dataset-size, and the DNN-MT method reaches up to $30\%$ higher spectral efficiency than the GIFP methods with different spatial and angular bin sizes. Also, Fig. \ref{Fig:DiffDatasets}\subref{Fig:_SEDE1} shows the performance of different beam alignment approaches with the minimal dataset, $\mathbb{D}^{\Xi}_{1\%}$, including only $700$ samples. As it can be seen, the DNN-EMT method has the best performance in terms of misalignment probability and ESE, regardless of the size of the candidate list. Following our expectation, by limiting the dataset size significantly, the performance gap between data-driven beam alignment methods and the HBS method is reduced. These results illustrates the trade off between accuracy and dataset size in DNN based methods: with large datasets, the DNN-ST model with a large number of trainable parameters leads to more accurate results than DNN-MT and DNN-EMT; when training data is scarce, however, the DNN-MT and DNN-EMT achieve better performance due to their much smaller number of trainable parameters.

\begin{figure}[t]
	\centering
	\subfloat[$\mathbb{D}^{\Xi}$ with  $70,000$ samples]{%
       \scalebox{0.899}{
       \includegraphics[width=0.45\textwidth, trim={0.68cm 0.6cm 2.3cm 1.5cm},clip]{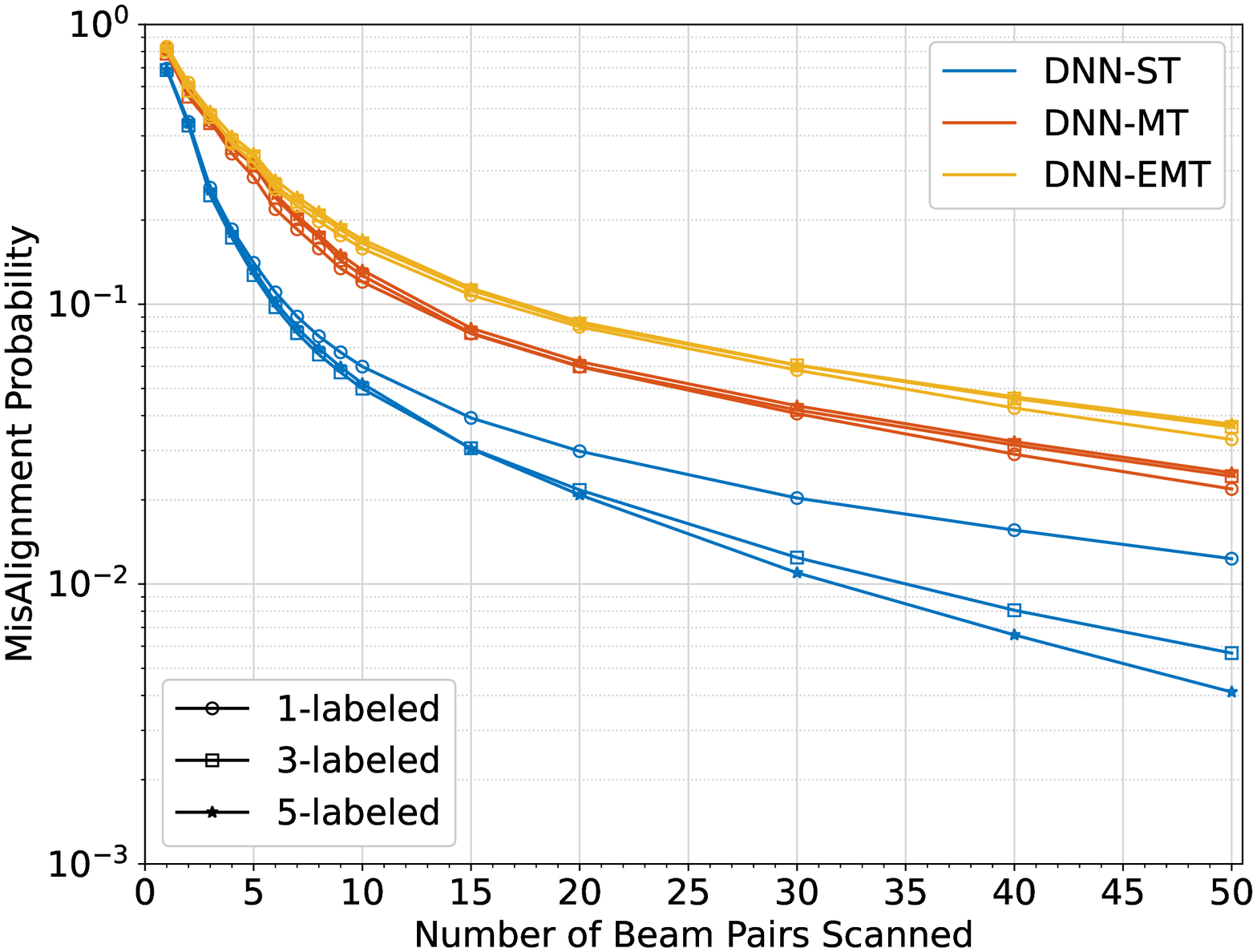}}}
    \hspace{2em}%
    \subfloat[$\mathbb{D}^{\Xi}_{1\%}$ with $700$ samples]{%
       \scalebox{0.899}{
       \includegraphics[width=0.45\textwidth, trim={0.68cm 0.6cm 2.3cm 1.5cm},clip]{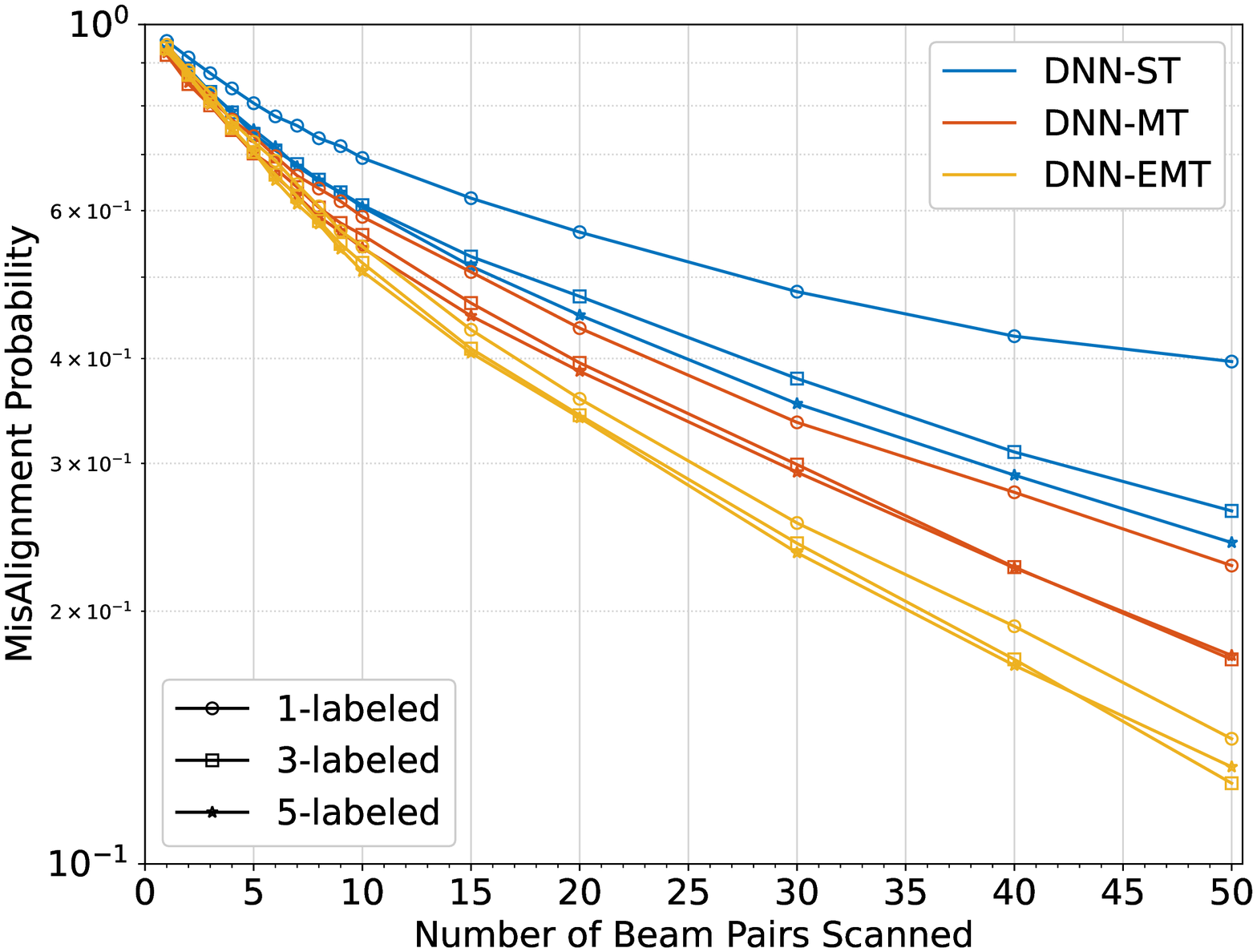}}}  
	\caption{Effects of the multi-labeling technique on the performance of the proposed neural network structures with large and small training datasets.}
	\label{Fig:MultiLabeling}
\end{figure}

Fig. \ref{Fig:MultiLabeling} shows the misalignment probability of the deep learning-based beam selection methods using the labels with $M$-hot encoding scheme. We use both $\mathbb{D}^{\Xi}$ and $\mathbb{D}^{\Xi}_{1\%}$ datasets to see the effects of the multi-labeling technique on the performance of the methods with different structures. As there are many trainable parameters in the DNN-ST structure, multi-labeling works like data augmentation to fill the performance gap because of insufficient training samples. However, the DNN-MT and DNN-EMT structures have significantly fewer trainable parameters. Thus, training these methods using multi-labeling with the large dataset $\mathbb{D}^{\Xi}$ does not show any advantage over using a single label. When the training examples are limited in $\mathbb{D}^{\Xi}_{1\%}$, however, the multi-labeling technique is helpful to have a lower misalignment probability. 
\begin{figure}[t]
	\centering
    \scalebox{0.3}{
		\includegraphics[trim={0.6cm 0.6cm 2.3cm 1.5cm},clip]{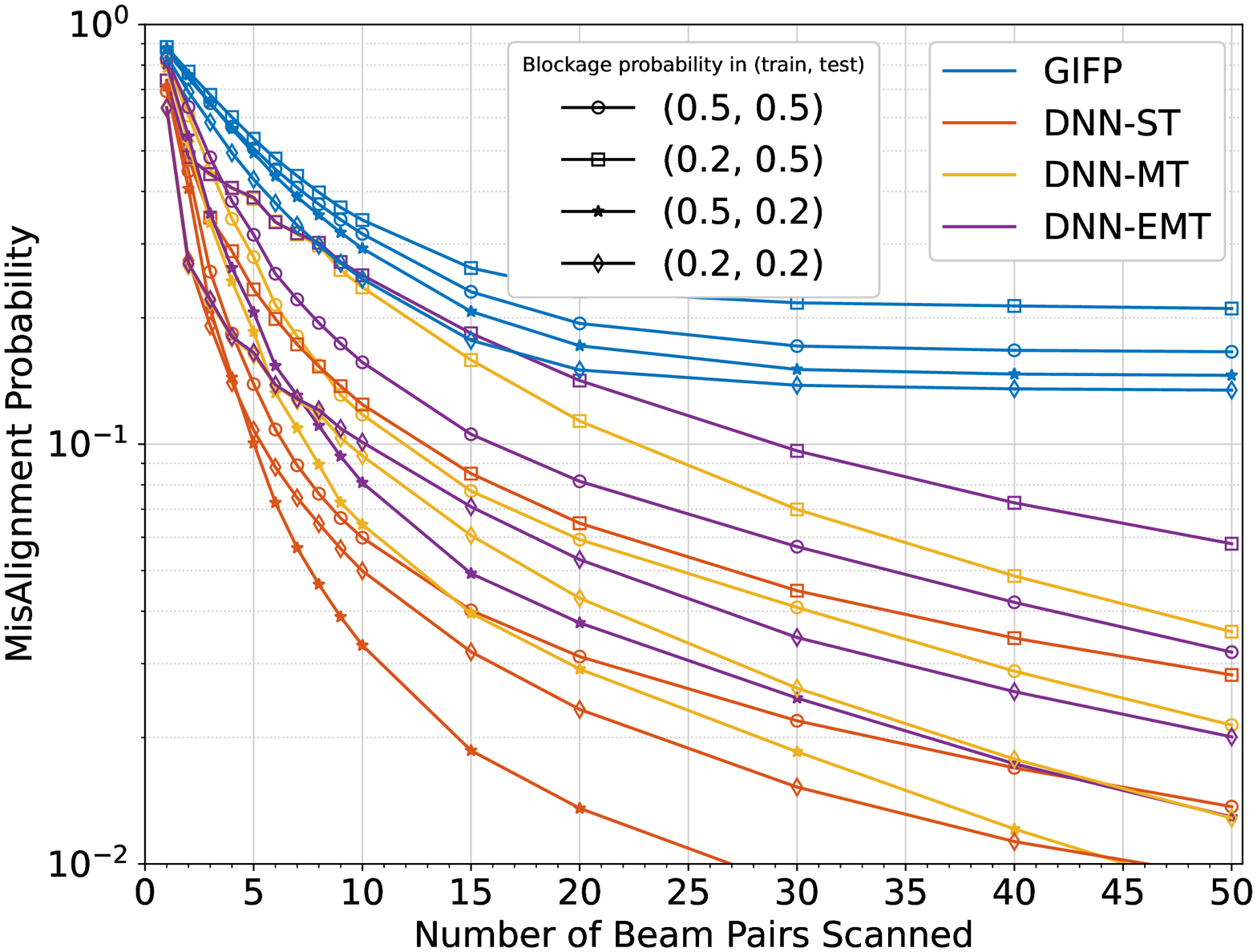}}
    \caption{The misalignment probability of different data-driven beam alignment methods with various LOS blockage probability in the training and test datasets.}
	\label{Fig:DiffBlockage}
\end{figure}

The robustness of the data-driven based beam alignment methods against different propagation properties in the training and test samples is evaluated in an experiment. We consider to have two datasets with $70,000$ samples, $\mathbb{D}^{\Xi}$ and $\mathbb{D}^{\Upsilon}$, with LOS blockage probability of $0.5$ and $0.2$, respectively. For the rest of simulations, we use SBS = $1$m and ABS = $45^{\circ}$ parameters for the GIFP method. Fig. \ref{Fig:DiffBlockage} shows the performance of the GIFP and deep learning-based methods when there is a match/mismatch of LOS blockage rate between the training and test samples. As it can be seen in Fig. \ref{Fig:DiffBlockage}, in the case where the LOS blockage rate is lower in the training set than in the test set, the network learns better the LOS direction but has fewer opportunities to learn NLOS directions. Thus, in comparison to the matched case when there is high LOS blockage probability in the training and test sets, the performance is degraded. Contrary, by training a neural network at a higher LOS blockage probability compared to the test samples, the network learns better the NLOS paths at the expense of a worse learning of the LOS one. Due to this, the network trained with matched LOS blockage probability performs better when few beam pairs are scanned. On the other hand, when more beam pairs are scanned there is a good chance to scan candidates for both LOS and NLOS paths. Thus, the network trained in a higher LOS blockage rate has a better performance compared to the matched case with large $N_b$. To summarize, the neural networks are robust to mismatch in the LOS blockage probability, and we see more robustness in the networks if trained in a high LOS blockage probability.

Although context information can reduce the overhead of initial beam alignment, in practical scenarios the position and orientation of the UT are subject to estimation errors. To account for this in our numerical assessment, we use a simple model of such inaccuracies. Consider $\boldsymbol{p}_\mathrm{UT}$ and $\boldsymbol{\psi}_\mathrm{UT}$ as the true position and orientation vector of the UT. By adding random perturbation to these vectors, we can model the measured position and orientation, $\boldsymbol{\hat{p}}_\mathrm{UT}$ and $\boldsymbol{\hat{\psi}}_\mathrm{UT}$, as
\begin{equation}
%\begin{split}
    \boldsymbol{\hat{p}}_\mathrm{UT} = \boldsymbol{p}_\mathrm{UT} + \boldsymbol{\epsilon}_p,
\end{equation}
\begin{equation}
    \boldsymbol{\hat{\psi}}_\mathrm{UT} = \boldsymbol{\psi}_\mathrm{UT} + \boldsymbol{\epsilon}_\psi,
%\end{split}
\end{equation}
where $\boldsymbol{\epsilon}_p$ and $\boldsymbol{\epsilon}_\psi$, respectively, denote the random perturbation of the position and orientation sensors on the UT. In line with the known statistics of the measuring error, the entries of the random perturbation vectors $\boldsymbol{\epsilon}_p$ and $\boldsymbol{\epsilon}_\psi$ are generated from independent Gaussian distributions with zero mean and variance $\sigma^2_p$ and $\sigma^2_\psi$, respectively.

\begin{figure}[t]
	\centering
	\subfloat[Inaccurate position\label{Fig:_SEP}]{%
       \scalebox{0.899}{
       \includegraphics[width=0.45\textwidth, trim={0.68cm 0.6cm 2.3cm 1.5cm},clip]{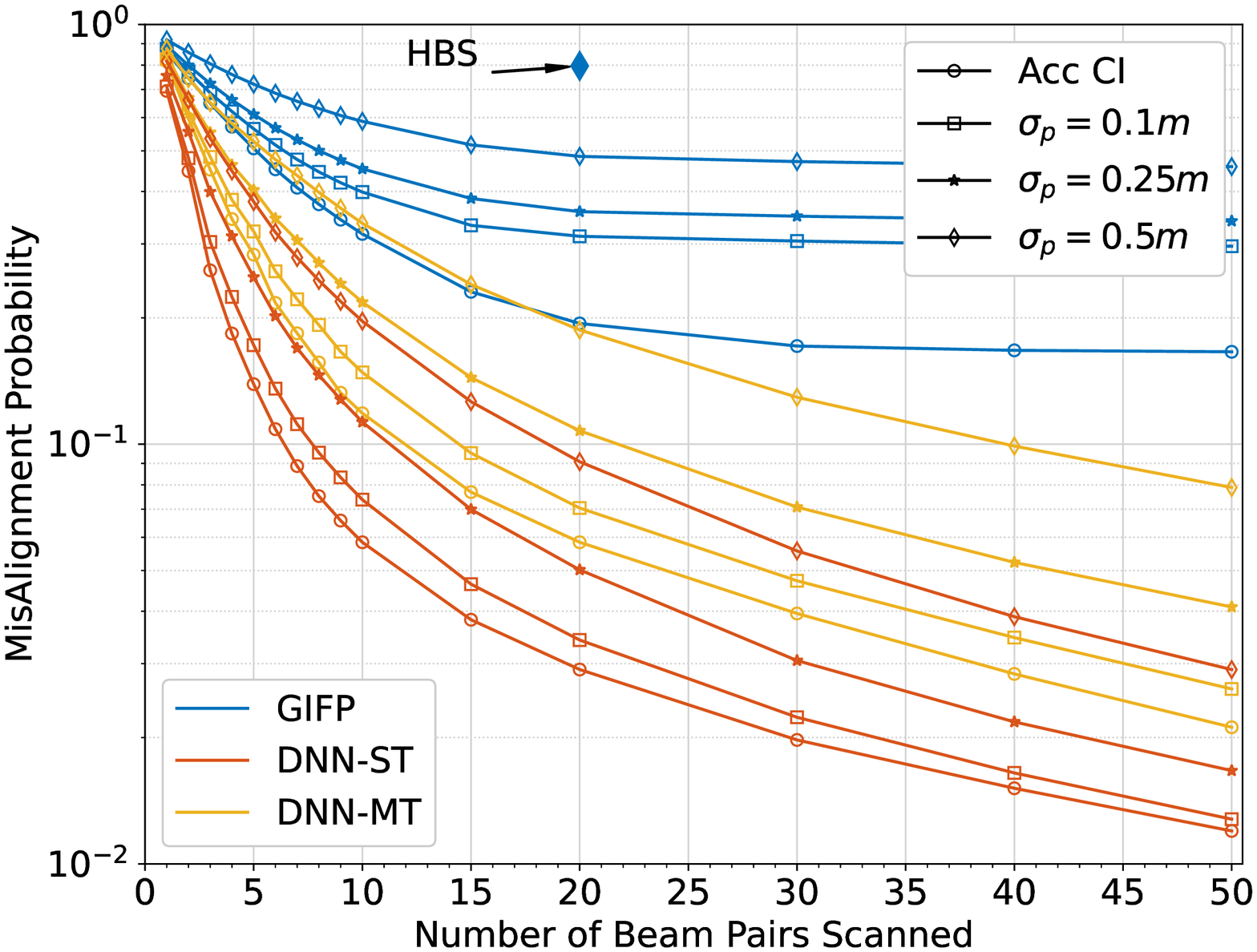}
       \hspace{2em}%
       \includegraphics[width=0.45\textwidth, trim={0.8cm 0.6cm 2.3cm 1.5cm},clip]{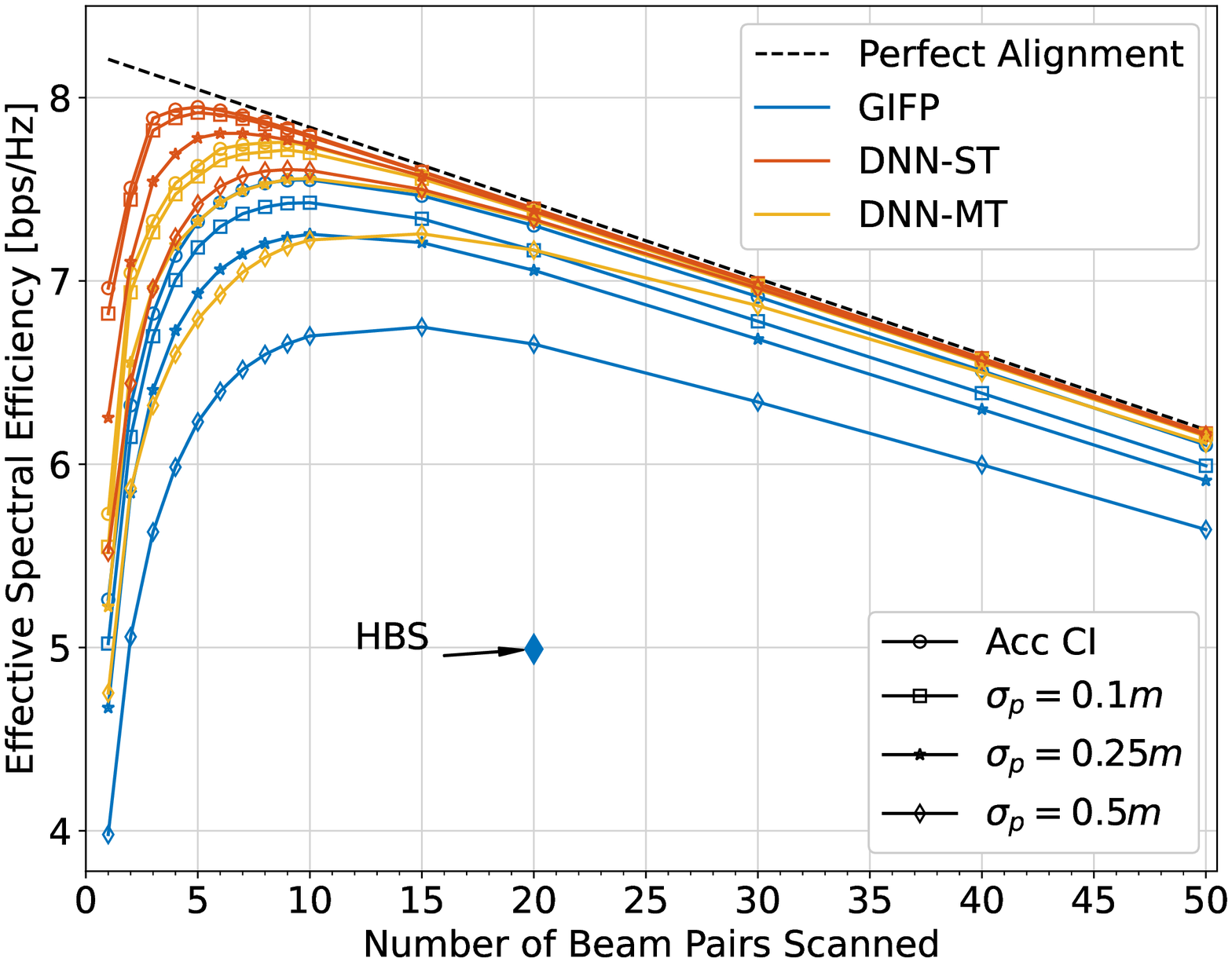}}}
    \\
    \subfloat[Inaccurate orientation\label{Fig:_SEO}]{%
       \scalebox{0.899}{
       \includegraphics[width=0.45\textwidth, trim={0.68cm 0.6cm 2.3cm 1.5cm},clip]{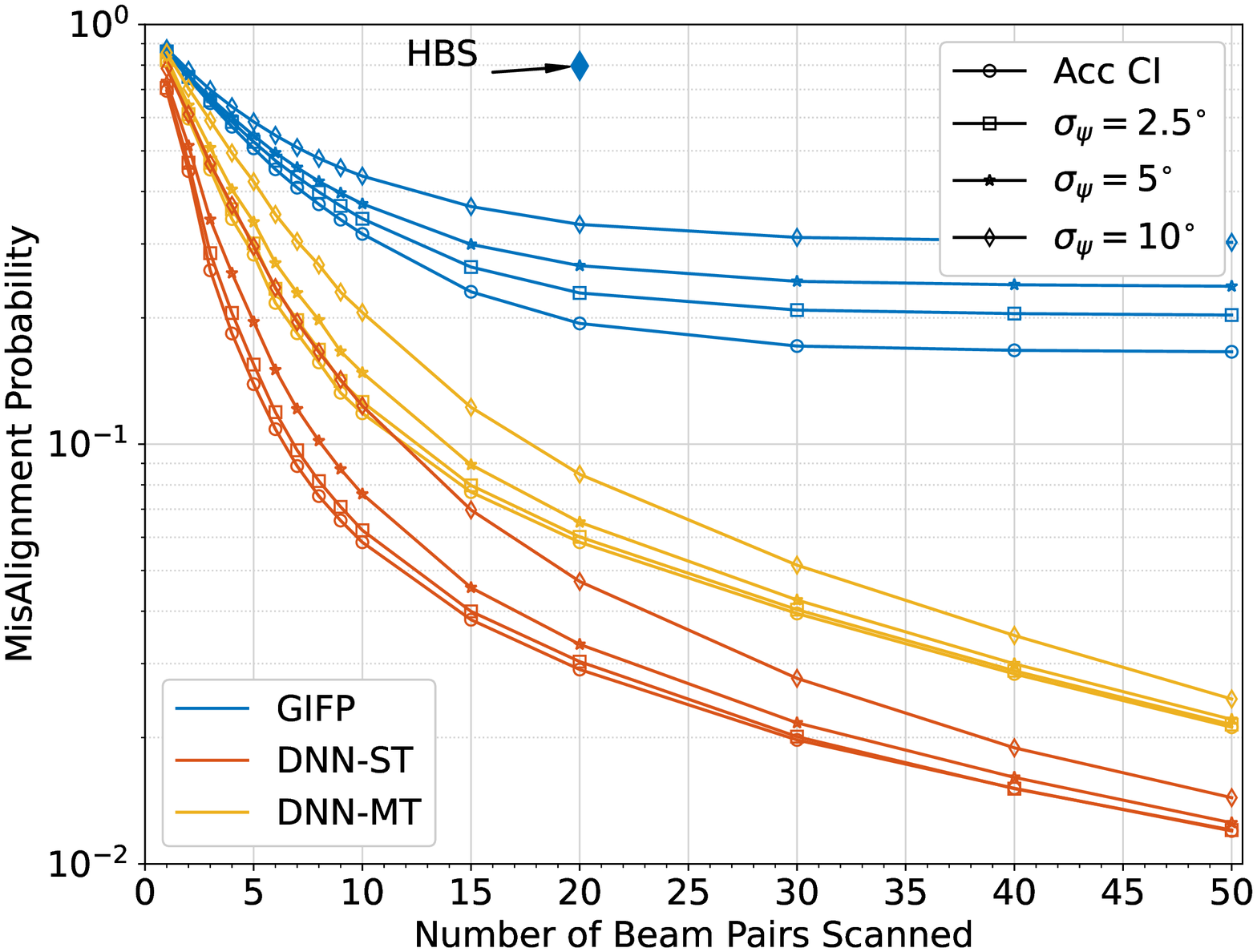}
       \hspace{2em}%
       \includegraphics[width=0.45\textwidth, trim={0.8cm 0.6cm 2.3cm 1.5cm},clip]{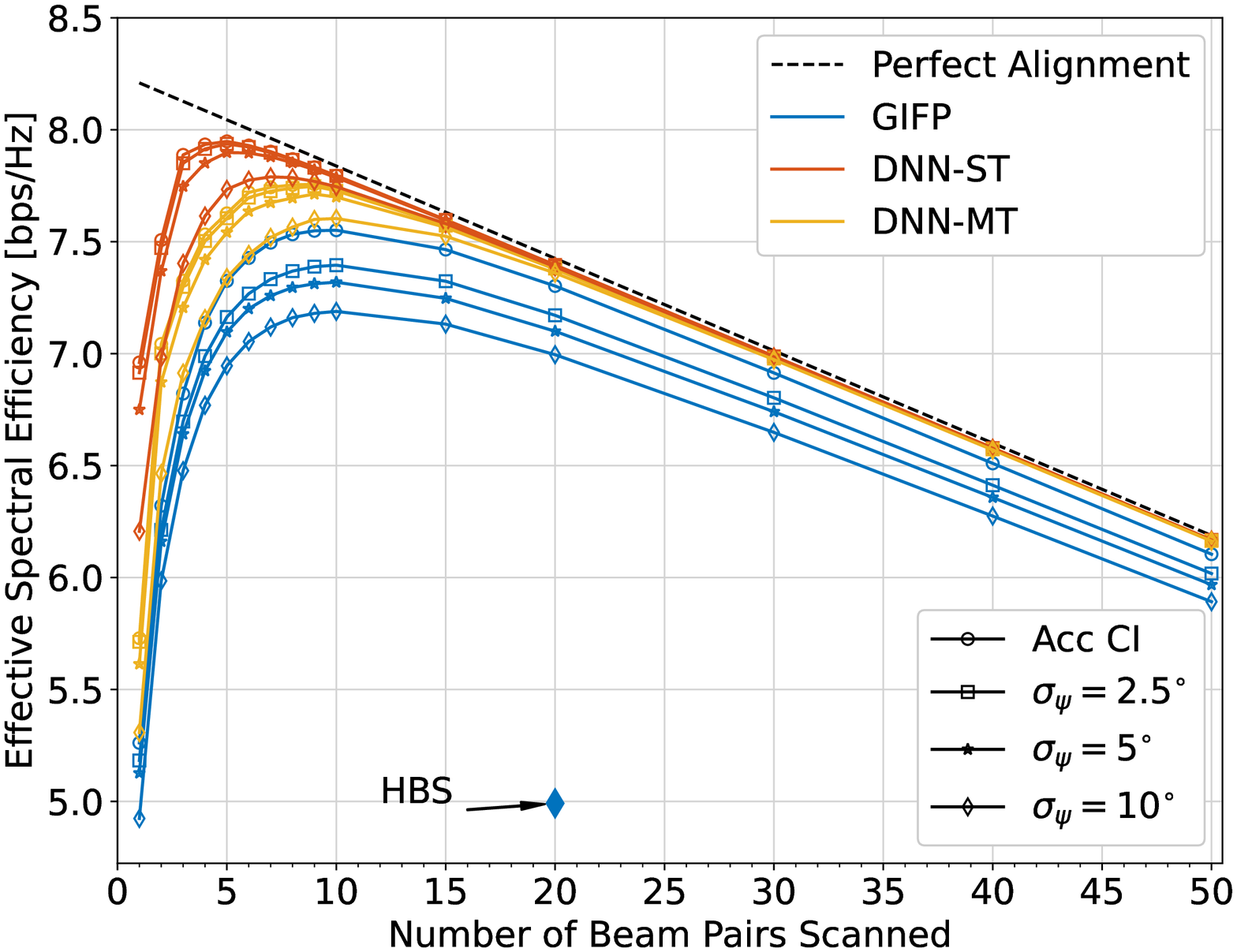}}}
    \\
    \subfloat[Inaccurate position and orientation\label{Fig:_SEPO}]{%
       \scalebox{0.899}{
       \includegraphics[width=0.45\textwidth, trim={0.68cm 0.6cm 2.3cm 1.5cm},clip]{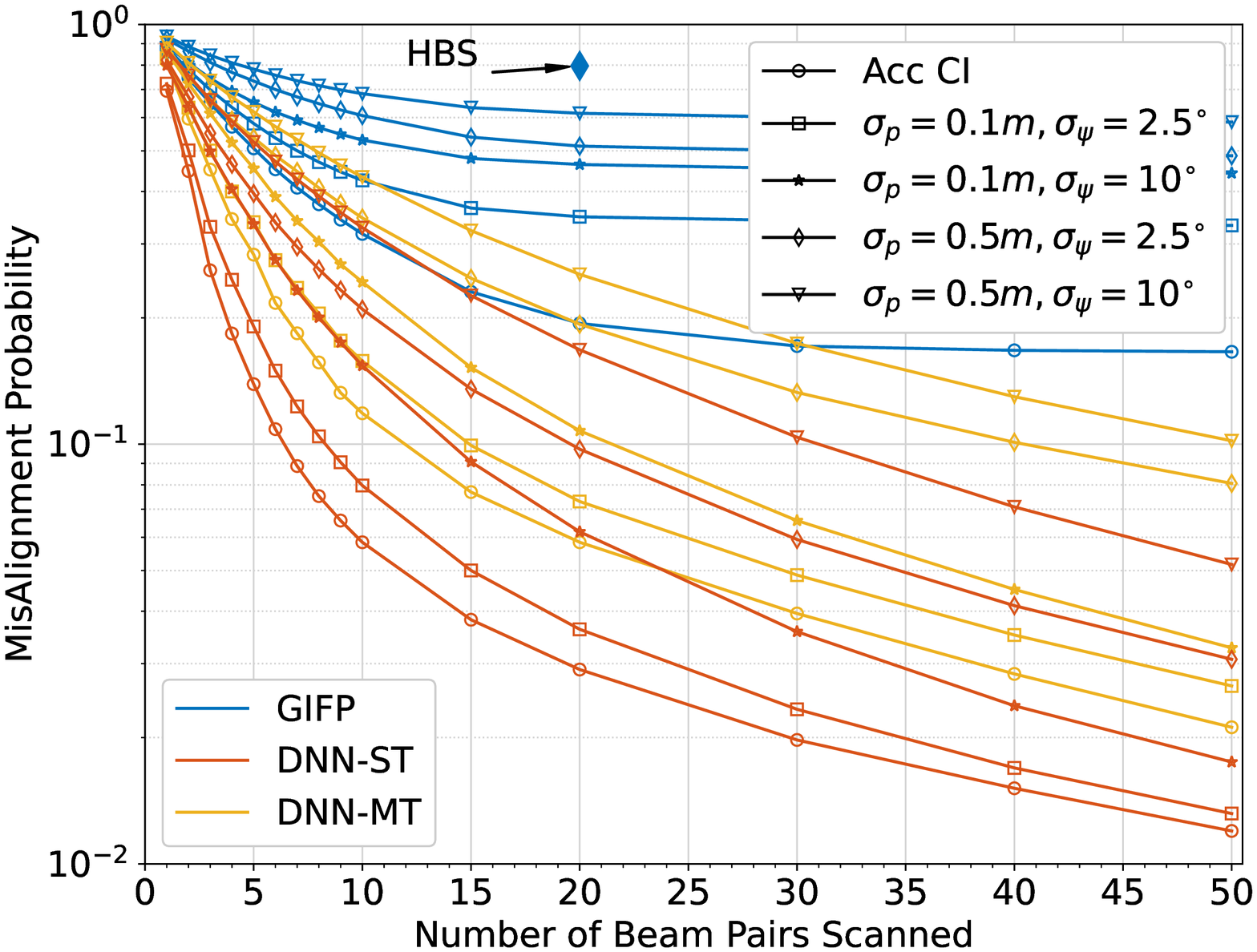}
       \hspace{2em}%
       \includegraphics[width=0.45\textwidth, trim={0.8cm 0.6cm 2.3cm 1.5cm},clip]{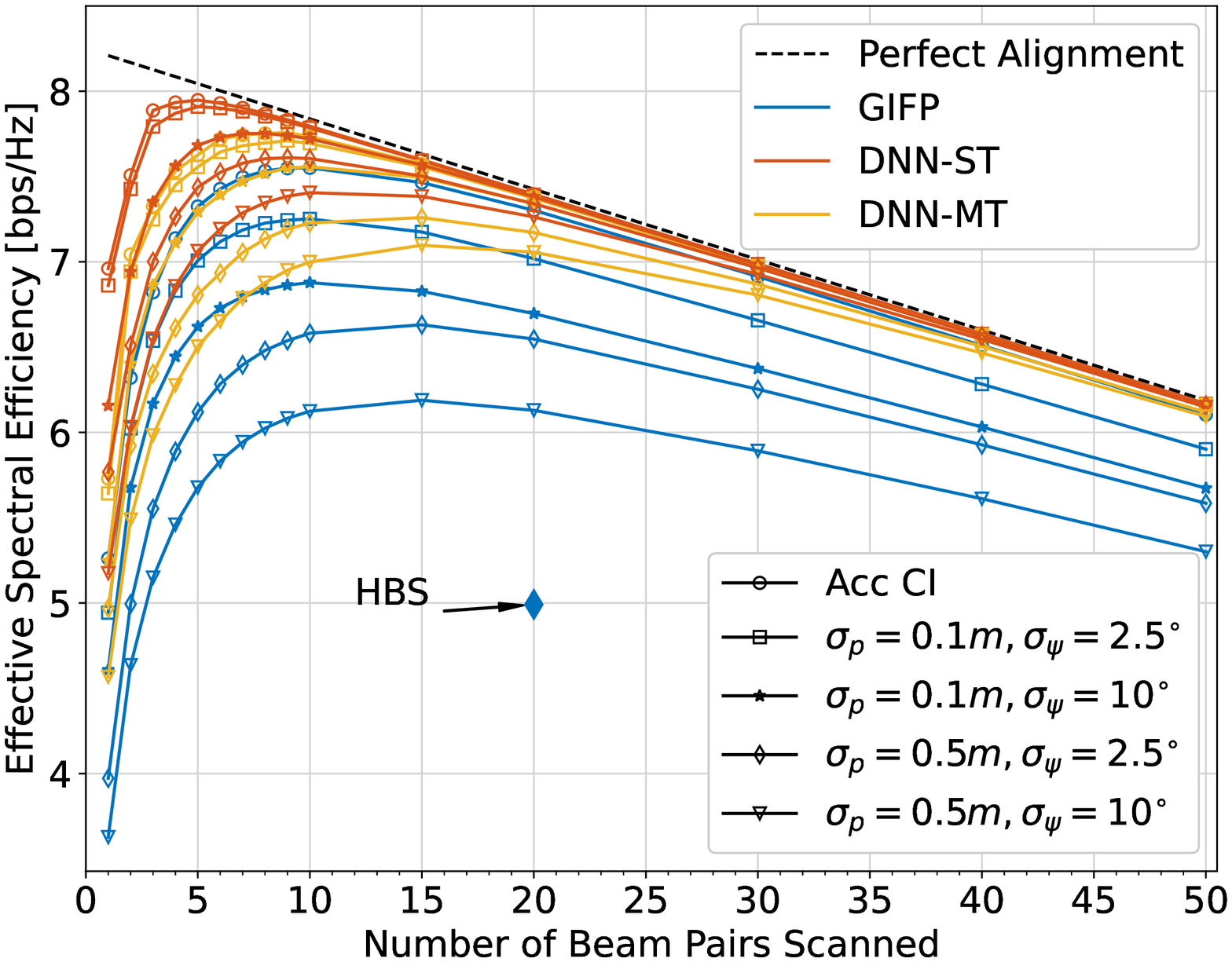}}}
	\caption{Misalignment probability and effective spectral efficiency with inaccurate position and orientation information.}
	\label{Fig:InaccCI}
\end{figure}

In Fig. \ref{Fig:InaccCI}, the performance of the GIFP, DNN-ST, and DNN-MT methods including inaccuracies in the position and/or orientation information of the UT device are shown. The dataset with $70,000$ samples are used, and the estimation error is included in both the train and test sets. When applying perturbation only on the position information, we see a slight degradation with $\sigma_p = 0.1m$. This level of accuracy can be obtained with, e.g., ultra wide-band (UWB) positioning \cite{zhang_real-time_2010}. As shown in Fig. \ref{Fig:InaccCI}\subref{Fig:_SEP}, by increasing $\sigma_p$ to $0.5m$, both the GIFP and deep learning-based methods have performance loss, respectively, around $10\%$ and $5\%$ in the maximum of achievable ESE. The sensitivity of data-driven based beam alignment methods to position inaccuracy depends on the dimensions of the environment. Thus, we expect to see less performance degradation due to inaccurate positions in big rooms or outdoor scenarios. Fig. \ref{Fig:InaccCI}\subref{Fig:_SEO} shows the performance of different beam alignment methods by feeding inaccurate orientation information. All the data-driven approaches offer good robustness to the different amounts of orientation perturbations. When we apply the perturbation model on both position and orientation information, as we expected, more degradation in their performance can be seen in Fig. \ref{Fig:InaccCI}\subref{Fig:_SEPO}. At all the considered $\sigma_p$ and $\sigma_o$ cases, the data-driven based methods show better performance than the HBS method. Thus, the data-driven methods still leverage inaccurate context information to reduce the latency of the beam alignment procedure and reach higher spectral efficiency. By looking at the ESE curves, it can be seen that the DNN-ST method has the least sensitivity to inaccurate location and orientation information. This shows that, during the training process, the DNN is able to learn how to deal with different degrees of CI uncertainty, leading to a robust performance against these inaccuracies.

\iffalse
Inaccurate CI causes performance degradation in data-driven beam alignment methods. To make neural networks robust against the inaccuracy of inputs, we use the data augmentation technique. In this technique, we replicate the inaccurate samples, $\mathbb{D}^{\Xi}$, $k$ times by adding synthetic Gaussian noise to the inaccurate samples. Data augmentation helps to feed the neural network more samples with more variations. The replicated dataset, $\mathbb{D}^{R}$, maybe emphasized as
\begin{equation}
    \mathbb{D}^{R} = \{ \mathbb{D}^{\Xi}, \mathbb{D}^{\Xi}_{n_2} , \dots, \mathbb{D}^{\Xi}_{n_k}\}
\end{equation}
where $\mathbb{D}^{\Xi}_{n_i}, i=2, \dots, k$ are the synthetic noisy version of $\mathbb{D}^{\Xi}$. For each of them, we added synthetic noise generated from a zero-mean Gaussian distribution with the same variance of random perturbation of the sensors.
\fi

\section{Conclusions}\label{Sec:Conc}
In this work, we have shown that position and orientation of UT as a kind of context information can play an important role in the initial beam alignment procedure of mmWave links in pedestrians applications. We have presented a deep-learning based approach for beam recommendation based solely on the position and orientation of the UT. Three neural network structures are designed for the task: a standard feed forward network coined DNN-ST, and two alternative and simplified designs. The simplified structures, coined DNN-MT and DNN-EMT, are inspired by multitask networks and separate the beamforming tasks at AP and UT, as well as the horizontal and vertical beam directions in the latter design. This results in a significant reduction of trainable parameters and size of the DNN model. 

Our extensive performance assessment based on ray-traced channel responses reveals that the DNN-ST design shows excellent performance when enough training samples are in the dataset. However, the DNN-MT and DNN-EMT designs have less computational cost in the training and evaluation phase, and these networks perform better with small training datasets. The proposed beam selection methods outperform the GIFP method as a CI-based lookup table approach and the DEACT method as a CI-agnostic hierarchical beam search solution in terms of latency and spectral efficiency. Our proposed methods show low sensitivity to changes in the propagation properties in the evaluation compared to the data collection time as well as displaying high robustness against measurement inaccuracies in the position and orientation of the UT. 

%As potential future works, an extension of this work to unmanned aerial vehicle (UAV) applications, considering multiple panels at the UT, and serving multiple users using hybrid analog-digital architecture are interesting directions.

\iffalse
\begin{itemize}
    \item CI-based beam alignment can be used in pedestrian applications by including the orientation information of UT
    \item The proposed deep learning-based method outperforms the GIFP and HBS methods
    \item Separation of beamforming in AP and UT reduces significantly the total number of trainable parameters, which leads to better performance with limited training datasets
    \item Besides the transfer learning technique, Multi Task structure is another approach to deal with limited training datasets. Combination of these approaches as an interesting idea can lead to even more gain.
    \item deep learning-based beam selection methods are robust to changing the probability of LOS blockage in the training and test datasets
    \item Inaccuracy in context information degrades the performance of CI-based methods, but with reasonable perturbations, they still work better than HBS
    \item Future studies: Multi-user / multi panel / UAV
    \item At the end, it makes totally sense to include position and orientation information in the beam alignment procedure of mmWave links.
\end{itemize}
\fi
% \section*{Acknowledgment}

% This work is supported by the Danish Council for Independent Research, grant no. DFF 8022-00371B.

\bibliographystyle{IEEEtran} % We choose the &quot;plain&quot; reference style
\bibliography{BeamAlignment_Ref} % Entries are in the &quot;refs.bib&quot; file</code></pre>

\end{document}